\documentclass[preprint2]{aastex631}
\usepackage[utf8]{inputenc}
\usepackage{amsmath}
\usepackage{ulem}
\usepackage{bm}
\usepackage[caption=false]{subfig}

\begin{document}

\shorttitle{NIR PLZ relations for RR Lyrae Stars Based on Gaia EDR3 Parallaxes}
\shortauthors{Zgirski et al.}

\title{New Near-Infrared Period-Luminosity-Metallicity Relations for Galactic RR Lyrae Stars Based on Gaia EDR3 Parallaxes.}

\author{Bartłomiej Zgirski}
\affil{Nicolaus Copernicus Astronomical Center, Polish Academy of Sciences, Bartycka 18, 00-716 Warszawa, Poland}
\email{bzgirski@camk.edu.pl}
\author{Grzegorz Pietrzyński}
\affil{Nicolaus Copernicus Astronomical Center, Polish Academy of Sciences, Bartycka 18, 00-716 Warszawa, Poland}
\affil{Universidad de Concepción, Departamento de Astronomía, Casilla 160-C, Concepción, Chile}
\author{Marek Górski}
\affil{Nicolaus Copernicus Astronomical Center, Polish Academy of Sciences, Bartycka 18, 00-716 Warszawa, Poland}
\author{Wolfgang Gieren}
\affil{Universidad de Concepción, Departamento de Astronomía, Casilla 160-C, Concepción, Chile}
\author{Piotr Wielgórski}
\affil{Nicolaus Copernicus Astronomical Center, Polish Academy of Sciences, Bartycka 18, 00-716 Warszawa, Poland}
\author{Paulina Karczmarek}
\affil{Universidad de Concepción, Departamento de Astronomía, Casilla 160-C, Concepción, Chile}
\author{Gergely Hajdu}
\affil{Nicolaus Copernicus Astronomical Center, Polish Academy of Sciences, Bartycka 18, 00-716 Warszawa, Poland}
\author{Megan Lewis}
\affil{Nicolaus Copernicus Astronomical Center, Polish Academy of Sciences, Bartycka 18, 00-716 Warszawa, Poland}
\author{Rolf Chini}
\affil{Nicolaus Copernicus Astronomical Center, Polish Academy of Sciences, Bartycka 18, 00-716 Warszawa, Poland}
\affil{Ruhr University Bochum, Faculty of Physics and Astronomy, Astronomical Institute (AIRUB), 44780 Bochum, Germany}
\affil{Universidad Católica del Norte, Instituto de Astronomía, Avenida Angamos 0610, Antofagasta, Chile}
\author{Dariusz Graczyk}
\affil{Nicolaus Copernicus Astronomical Center, Polish Academy of Sciences, Rabiańska 8, 87-100, Toruń, Poland}
\author{Mikołaj Kałuszyński}
\affil{Nicolaus Copernicus Astronomical Center, Polish Academy of Sciences, Bartycka 18, 00-716 Warszawa, Poland}
\author{Weronika Narloch}
\affil{Universidad de Concepción, Departamento de Astronomía, Casilla 160-C, Concepción, Chile}
\author{Bogumił Pilecki}
\affil{Nicolaus Copernicus Astronomical Center, Polish Academy of Sciences, Bartycka 18, 00-716 Warszawa, Poland}
\author{Gonzalo Rojas García}
\author{Ksenia Suchomska}
\affil{Nicolaus Copernicus Astronomical Center, Polish Academy of Sciences, Bartycka 18, 00-716 Warszawa, Poland}
\author{Mónica Taormina}
\affil{Nicolaus Copernicus Astronomical Center, Polish Academy of Sciences, Bartycka 18, 00-716 Warszawa, Poland}

\begin{abstract}
We present new period-luminosity and period-luminosity-metallicity relations for Galactic RR Lyrae stars based on a sample of 28 pulsators located at distances up to $1.5$ kpc from the Sun. Near-infrared photometry was obtained at the Cerro Armazones Observatory and parallaxes were taken from the Gaia Early Data Release 3. Relations were determined for the 2MASS $JHK_s$ bands and the $W_{JK}$ Wesenheit index. We compare our results with other calibrations available in the literature and obtain very good agreement with the photometry of RR Lyraes from the Large Magellanic Cloud anchored using the distance to the Cloud, which based on detached eclipsing binaries. We find that the dependence of absolute magnitudes on metallicity of $0.070\pm 0.042$ mag/dex ($J-$ band) to $0.087 \pm 0.031$ mag/dex ($W_{JK}$ index) for the population of fundamental pulsators (RRab) that is in agreement with previously published phenomenological works. We perform a refined determination of distance to the LMC based on our new calibration and photometry from \cite{SZEWCZYK-LMC}. We study the dependence of the fitted parameters of fiducial relations and the LMC distance on the systematic parallax offset.
\end{abstract}


\section{Introduction}
RR Lyrae (RRL) stars are one of the most studied types of variable stars (e.g., \citealt{PULS_STARS}, \citealt{BHARDWAJ}). They are old (with ages of above 10 Gyr), metal-poor, helium-burning stars from the intersection of the instability strip and the horizontal branch. These tracers of Population II may be found in the bulge, halo, and thick disk of the Galaxy as well as in globular clusters. Their pulsation periods are typically 0.2 - 1  day.

RRL stars, discovered at the end of the 19th century, have been used as distance indicators since \cite{SHAPLEY} who first used them to determine distances to a number of globular clusters in the Milky Way, thus allowing for the estimation of the size of the Galaxy. Even though RRL stars are fainter than classical Cepheids, they serve to determine distances to old populations such as dwarf galaxies in the neighborhood of the Milky Way (e.g. the distances to Sculptor dSph, Carina, and Fornax galaxies by \citealt{SCULPTOR}, \citealt{CARINA}, \citealt{FORNAX}). They are also useful in the structure mapping of stellar systems (see e.g. \citealt{JACYSZYN} for such a mapping of the Magellanic system). They provide an independent way of testing different long-range distance determination methods, such as the tip of the red giant branch (TRGB), that are crucial in the calibration of the Hubble constant (e.g. \citealt{Beaton}).

Especially the near-infrared (NIR) period-luminosity (PL) relations for RRL stars serve as accurate distance determination tools. Pulsation amplitudes are smaller in the NIR domain as compared to the optical; thus accurate mean magnitudes of these stars can be obtained based merely on a few photometric epochs. Additionally, the influence of reddening, an important systematic factor in distance determinations, is considerably decreased in the NIR. The width of the instability strip in the NIR is smaller compared to optical bands. These factors contribute to the relatively small scatter of residuals in period-luminosity relations at longer wavelengths. The first PL relations for RRL stars in the NIR $K-$ band were established by \cite{LONGMORE}.

In parallel, the influence of metallicity on the luminosities of RR Lyrae stars has been studied. \citeauthor{SANDAGE1958} (\citeyear{SANDAGE1958}, \citeyear{SANDAGE1981}) showed that RR Lyrae stars from the metal-poor Oosterhoff-II group are more luminous than their more metal-rich counterparts, which led to the establishment of a linear relation between the $V-$ band absolute magnitudes of RRL stars and their [Fe/H] metallicities \citep{SANDAGE1990}. 

The accurate calibration of NIR period-luminosity-metallicity (PLZ) relations for RRL stars became an important task for the community and has been studied from both a theoretical (e.g. \citealt{Bono2001}, \citealt{Bono2003}, \citealt{CATELAN-PLZ}, \citealt{BRAGA}, \citealt{MARCONI}), and an empirical perspective (e.g. \citealt{Sollima2006}, \citeyear{SOLLIMA}, \citealt{MURAVEVA2015}, \citealt{MURAVEVA_GDR2}, \citealt{NEELEY}, \citealt{VMC}, \citealt{BPLZ}). Such relations are usually presented for the mixed population of fundamental (RRab) and the first-overtone (RRc) pulsators where the fundamentalization factor for the periods of RRc stars is taken from \cite{Iben0127}: $log P_{ab}=log P_c + 0.127$.

Calibrations of PLZ relations can be anchored to globular clusters or the Large  Magellanic Cloud (LMC) where individual distances for all stars from such samples have been assumed (e.g. \citealt{MURAVEVA2015}, \citealt{VMC}). Some calibrations were tied to parallaxes of single field stars (e.g. \citealt{Sollima2006}, \citeyear{SOLLIMA} with a zero point of the PLZ relation from the Hubble Space Telescope - HST - parallax of the protoplast star RR Lyr; \citealt{MURAVEVA2015} with a zero point from HST parallaxes of 4 RRL stars).

The advent of very precise parallaxes for field RRL stars from the solar neighborhood from the Gaia space mission (\citealt{GAIAEDR3} - Gaia Early Data Release 3 - EDR3) allowed for the determination of absolute magnitudes of these stars with precisions better than previously available. It also provides an alternative, independent method of verification of the zero points for calibration of PLZ relations for RRL stars.

In our paper, we take advantage of the Gaia parallaxes of nearby (distances up to 1.5 kpc) RRL stars, NIR photometry gathered especially for this project using the IRIS instrument \citep{HODAPP} at the Cerro Armazones Observatory \citep{RAMOLLA}, and metallicities reported by \cite{CRESTANI} in order to establish fiducial PL and PLZ relations.
This paper is the second in a series devoted to the calibration of NIR PL relations for pulsating stars from the solar neighborhood based on the photometry acquired at the Cerro Armazones Observatory, with the first paper \citep{WIELGOR} being devoted to Type II Cepheids.

\begin{figure*}
    \centering
    \plotone{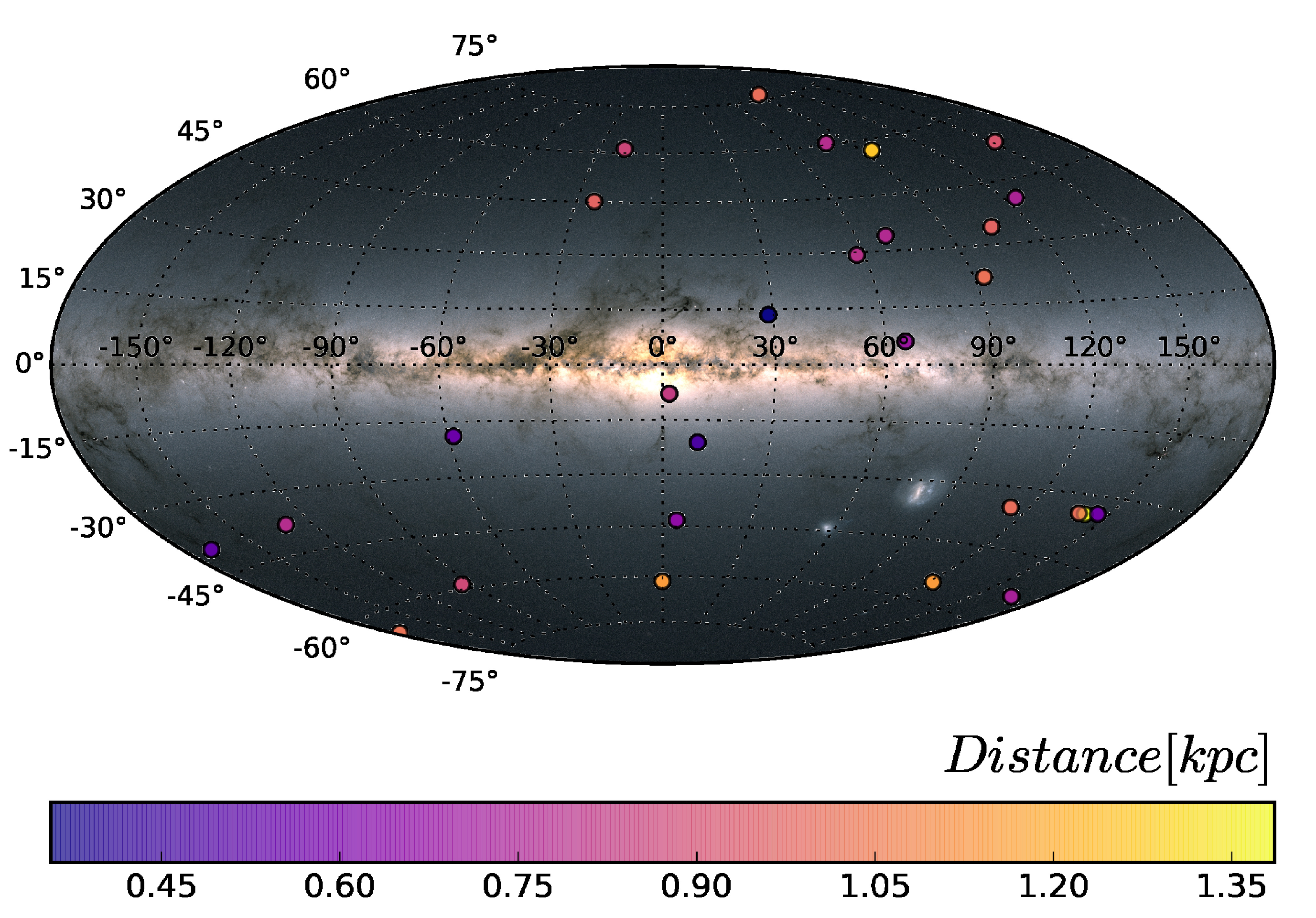}
    \caption{Location of 28 RR Lyrae stars in the Galactic coordinate system used to establish PL and PLZ relations; their distances, from Gaia EDR3, are indicated by colors. Background image ’The colour of the sky from Gaia’s Early Data Release 3’: \url{https://www.esa.int/ESA_Multimedia/Images/2020/12/The_colour_of_the_sky_from_Gaia_s_Early_Data_Release_3.}    \label{fig:GAIA_projection}}
\end{figure*}

\section{Data}
Between February 2017 and March 2020, we gathered NIR photometry of 28  RR Lyrae stars at distances up to 1.5 kpc from the Sun using the 0.8 m telescope of the \textit{Infra-Red Imaging  Survey} (IRIS) equipped with a HAWAII-1 NIR camera (\citealt{HODAPP}, \citealt{WATERMANN}), and $JHK_s$ filters similar to their counterparts from the 2MASS system \citep{2MASS}. The scientific frames were calibrated using an automated pipeline \citep{WATERMANN} based on IRAF \citep{IRAF}, SExtractor \citep{SEXTRACTOR}, and SCAMP \citep{SCAMP}. We performed aperture photometric measurements using a custom photometric pipeline based on the Python Astropy library \citep{ASTROPY} and DAOPHOT photometric package \citep{STETSON}. The custom pipeline also allowed for the instant standardization of the photometry onto the 2MASS system based on comparison stars whose photometry has been taken from the catalog of \cite{2MASSCAT}. The usual precision of the instrumental photometry was $0.01 - 0.02$\,mag. The accuracy of the overall photometric zero point, calculated as an error on the mean of differences between the catalog and independently derived magnitudes of constant control stars, is estimated at 0.002\,mag  \citep{WIELGOR}, and the mean shift between the standardized and 2MASS catalog’s zero points equal to zero. Figure \ref{fig:GAIA_projection} presents the location of RRL stars from our sample in the celestial sphere.

Pulsation periods ($P$) of RR Lyrae stars from our sample were taken from  the \textit{International Variable Star Index}\footnote{\url{https://www.aavso.org/vsx/}}. In the case of PL relations for the mixed population of RRab+RRc stars, we needed to apply shifts of logarithms of periods for RRc stars ($\Delta \log P$). Instead of utilizing the value of $\Delta \log P=+0.127$ \citep{Iben0127}, commonly used in the literature, we have independently fitted $\Delta \log P$ values. The main motivation for this was the suboptimal alignment of RRab and RRc stars in global PL relations for our sample when applying the standard fundamentalization value. See the subsection 4.2 for further elaboration on this topic.

We have determined mean stellar magnitudes by fitting a second-order Fourier series to the flux. Statistical errors of mean magnitudes were calculated as the mean uncertainty for all epochs divided by the square root of the number of epochs for a given object. Errors smaller than 0.005\,mag were rejected, and instead fixed to this value due to the uncertainty associated with the choice of the order of the Fourier series. The contribution of the photometric errors to the final uncertainties of the derived parameters of the PL and PLZ relations is much smaller than the components related to uncertainties of stellar parallaxes. 
In the case of 5 stars (DX Del, RR Leo, RU Psc, RY Col, UU Vir - only $J-$ band) less than 5 photometric epochs were obtained. We derived mean magnitudes of these stars by averaging fluxes. Following the reasoning presented by \cite{ScowcroftA} and \cite{NeeleyA}, uncertainties of mean magnitudes for randomly and sparsely covered light curves are dominated by the scatter of magnitudes. The largest amplitudes derived for well-covered light curves from our sample are $A_{J, max}=0.48$ mag, $A_{H, max}=0.39$ mag, and $A_{K, max}=0.35$ mag for the $J-$, $H-$, and the $K-$ band, respectively. Taking into account the conservative assumption of the uniform distribution of magnitudes, we obtain a single-measurement scatter of $\sigma=A/\sqrt{12}$, which yields the corresponding $\sigma_J=0.14$ mag, $\sigma_H=0.11$ mag, and $\sigma_K=0.10$ mag. It is further divided by the square root of the number of epochs in order to obtain uncertainties of mean magnitudes.
For two stars (HY Com \& SS Leo) only single comparison stars were used to standardize the photometry.
Mean magnitude errors were adopted as 0.01\,mag  in these two cases. Table \ref{tab:av_mags} depicts mean magnitudes, pulsation periods, and color excesses for stars from our sample. Plots depicting light curves of RR Lyrae stars from our sample in the three NIR bands are presented in Appendix A.

We dereddened the photometry based on the $E(B-V)$ color excess values reported by \cite{SCHLAFLY}. Such values were integrated assuming the three-dimensional Milky Way model of \cite{M-W_MODEL}. Extinctions corresponding to the three NIR bands were calculated from $E(B-V)$ assuming ratios of total-to-selective extinctions based on \cite{Cardelli} and the reddening law $R_V=3.1$. Namely, we adopted $A_J/A_V=0.288$, $A_H/A_V=0.180$, and $A_{K_s}/A_V=0.117$.

\begin{table*}
    \centering
    \begin{tabular}{|c|c|c|c|c|c|c|}
    \hline
ID & type & $<J>$ & $<H>$ & $<K_s>$ & $P$ & $E(B-V)$ \\
  &  & [mag] & [mag] & [mag] & [days] & [mag] \\ \hline
AE Boo & RRc & $ 9.934 \pm 0.005 $ & $ 9.759 \pm 0.005 $ & $ 9.729 \pm 0.005 $ & 0.31489 & 0.023 \\
AN Ser & RRab & $ 10.061 \pm 0.005 $ & $ 9.877 \pm 0.005 $ & $ 9.786 \pm 0.005 $ & 0.52207 & 0.036 \\
BB Eri & RRab & $ 10.543 \pm 0.005 $ & $ 10.298 \pm 0.005 $ & $ 10.221 \pm 0.005 $ & 0.56991 & 0.043 \\
DX Del & RRab & $ 8.901 \pm 0.080 $ & $ 8.696 \pm 0.060 $ & $ 8.648 \pm 0.060 $ & 0.47262 & 0.079 \\
EV Psc & RRc & $ 9.854 \pm 0.005 $ & $ 9.662 \pm 0.005 $ & $ 9.630 \pm 0.005 $ & 0.30626 & 0.030 \\
FW Lup & RRab & $ 7.994 \pm 0.005 $ & - & $ 7.718 \pm 0.005 $ & 0.48417 & 0.062 \\
HY Com & RRc & $ 9.692 \pm 0.010 $ & $ 9.563 \pm 0.010 $ & $ 9.426 \pm 0.010 $ & 0.44859 & 0.024 \\
IK Hya & RRab & $ 9.088 \pm 0.008 $ & $ 8.851 \pm 0.005 $ & $ 8.781 \pm 0.005 $ & 0.65032 & 0.055 \\
MT Tel & RRc & $ 8.321 \pm 0.005 $ & $ 8.148 \pm 0.005 $ & $ 8.111 \pm 0.005 $ & 0.31690 & 0.034 \\
RR Leo & RRab & $ 10.081 \pm 0.080 $ & $ 9.795 \pm 0.060 $ & $ 9.768 \pm 0.060 $ & 0.45240 & 0.035 \\
RU Psc & RRc & $ 9.347 \pm 0.080 $ & $ 9.112 \pm 0.060 $ & $ 9.109 \pm 0.060 $ & 0.39038 & 0.039 \\
RV Cet & RRab & $ 9.975 \pm 0.006 $ & $ 9.774 \pm 0.007 $ & $ 9.672 \pm 0.005 $ & 0.62341 & 0.027 \\
RX Eri & RRab & $ 8.702 \pm 0.005 $ & $ 8.452 \pm 0.005 $ & $ 8.358 \pm 0.005 $ & 0.58725 & 0.053 \\
RY Col & RRab & $ 10.19 \pm 0.10 $ & $ 9.987 \pm 0.080 $ & $ 9.913 \pm 0.070 $ & 0.47884 & 0.025 \\
SS Leo & RRab & $ 10.212 \pm 0.010 $ & $ 9.967 \pm 0.010 $ & $ 9.924 \pm 0.010 $ & 0.62634 & 0.017 \\
SV Eri & RRab & $ 8.824 \pm 0.005 $ & $ 8.630 \pm 0.005 $ & $ 8.552 \pm 0.005 $ & 0.71388 & 0.078 \\
SX For & RRab & $ 10.163 \pm 0.005 $ & $ 9.965 \pm 0.005 $ & $ 9.856 \pm 0.005 $ & 0.60534 & 0.012 \\
T Sex & RRc & $ 9.347 \pm 0.005 $ & $ 9.201 \pm 0.005 $ & $ 9.149 \pm 0.005 $ & 0.32470 & 0.042 \\
U Lep & RRab & $ 9.725 \pm 0.005 $ & $ 9.556 \pm 0.005 $ & $ 9.493 \pm 0.005 $ & 0.58148 & 0.029 \\
UU Vir & RRab & $ 9.880 \pm 0.080 $ & $ 9.562 \pm 0.005 $ & $ 9.528 \pm 0.005 $ & 0.47561 & 0.016 \\
V467 Cen & RRab & $ 9.522 \pm 0.005 $ & $ 9.394 \pm 0.005 $ & $ 9.251 \pm 0.005 $ & 0.55140 & 0.050 \\
V675 Sgr & RRab & $ 9.281 \pm 0.005 $ & $ 9.034 \pm 0.005 $ & $ 8.989 \pm 0.005 $ & 0.64229 & 0.089 \\
V753 Cen & RRc & $ 9.769 \pm 0.005 $ & $ 9.664 \pm 0.005 $ & $ 9.608 \pm 0.005 $ & 0.22135 & 0.147 \\
V Ind & RRab & $ 9.114 \pm 0.005 $ & $ 8.928 \pm 0.005 $ & $ 8.875 \pm 0.005 $ & 0.47960 & 0.04 \\
WY Ant & RRab & $ 9.835 \pm 0.005 $ & $ 9.710 \pm 0.005 $ & $ 9.653 \pm 0.005 $ & 0.57434 & 0.055 \\
WZ Hya & RRab & $ 9.920 \pm 0.005 $ & $ 9.695 \pm 0.005 $ & $ 9.630 \pm 0.005 $ & 0.53772 & 0.069 \\
X Ari & RRab & $ 8.267 \pm 0.005 $ & $ 8.052 \pm 0.005 $ & $ 7.903 \pm 0.005 $ & 0.65118 & 0.158 \\
XZ Gru & RRab & $ 9.713 \pm 0.005 $ & $ 9.440 \pm 0.005 $ & $ 9.369 \pm 0.005 $ & 0.88310 & 0.010 \\
\hline
median & & & & & 0.52990 & 0.040 \\
\hline
    \end{tabular}
    \caption{Apparent (reddened) mean magnitudes of RR~Lyrae stars observed with IRIS in $JHK_s$ together with their pulsation periods (from AAVSO VSX). Extinctions were estimated using the Milky Way model by \cite{M-W_MODEL} and the extinction maps from \cite{SCHLAFLY}.}
    \label{tab:av_mags}
\end{table*}

\section{Determination of PL and PLZ relations}

Based on mean apparent magnitudes, absolute magnitudes needed for the determination of period-luminosity-(metallicity) relations were derived using Gaia EDR3 parallaxes relying on four different techniques:

\begin{itemize}
    \item Using parallaxes inserted directly into the definition of the distance modulus, yielding the absolute magnitude $M=m+5\log\varpi +5$ ($M$ is the absolute and $m$ is the apparent magnitude, $\varpi$ is the parallax in arcsec).
    \item{Using distances of \cite{B-J2021} derived from Gaia parallaxes using direction-dependent priors on distance (geometric distances), $M=m-5\log r +5$ ($r$ - distance in pc).}
    \item{As above, but using distance priors dependent on direction, colors, and apparent magnitudes of stars (photo-geometric distances of \citealt{B-J2021}}).
    \item{Using the \textit{Astrometry-Based Luminosity} \citep[ABL,][]{ABL}, a quantity that is directly proportional to the parallax, $a:=10^{0.2M}=\varpi 10^{\frac{m+5}{5}}$. While dealing with relations based on stellar parallaxes, the ABL is an asymptotically unbiased estimator. The higher the number of stars used for the fit the better the precision of the mean values.}
\end{itemize}

Parallaxes with the renormalized unit weight error for astrometry RUWE $>1.4$ and the level of asymmetry of a source in the Gaia image ipd\_gof\_harmonic\_amplitude$>0.1$ should be rejected in order to avoid resolved binaries \citep{RUWE}. No stars were rejected from our sample after applying these criteria. We used parallaxes from Gaia EDR3 \citep{GAIAEDR3} after applying corrections from \cite{LINDEGREN}. \cite{B-J2021} already took into account these corrections. 

We have independently determined PL and PLZ relations for $JHK_s$ bands and the $W_{JK}=K_s-0.69(J-K)$ Wesenheit index \citep{WESENHEIT}. Mean stellar magnitudes used for determining relations for $W_{JK}$ were not dereddened.

\subsection{Period-luminosity relations}
For all considered types of fits of parameters of PL relations, we fitted the unweighted relation between the absolute magnitude (or ABL) and period using the \textit{optimize.leastsq} function of SciPy \citep{Scipy}.

The relation between absolute magnitude and period takes the form of:
\begin{equation}
    M_{\lambda}(P')=a_{\lambda}(\log P'+\Delta \log P_{\lambda} \times FO) +b_{\lambda}
    \label{eq:PLR}
\end{equation}
where $a_{\lambda}$ (the slope), $b_{\lambda}$ (the intercept), and $\Delta \log P_{\lambda}$ (the shift of $\log P$ for RRc stars) are fitted parameters for a given band $\lambda$. Pulsation mode is denoted by $FO$ and it takes the value of 1 for RRc and 0 for RRab stars. The logarithm of the pivot period value from $\log  P'=\log P-\log P_0$ has been set to $\log P_0=-0.25$ for relations for RRab and RRab+RRc stars. In the case of relations for only RRc stars, $\log P_0=-0.45$ in order to minimize uncertainties of intercept values.

In the case of the ABL, the fitted relation is as follows:
\begin{equation}
    ABL_{\lambda}(P')=10^{0.2\left[a_{\lambda}\left(\log P'+\Delta \log P_{\lambda} \times FO\right)+b_{\lambda}\right]}
\end{equation}

\newcommand\mc{\multicolumn}
\begin{table*}
\footnotesize
\centering
    \begin{tabular}{|l|l||c|c|c||c|c|c|}
    \hline
       \mc{2}{|l||}{PL relations} &\mc{3}{c||}{parallax}&\mc{3}{c|}{geometric distance}\\
       \hline
        \mc{1}{|l|}{band}&\mc{1}{l||}{population}& \mc{1}{c|}{$a$} & \mc{1}{c|}{$b$} &\mc{1}{c||}{$\Delta \log P$} &  \mc{1}{c|}{$a$} & \mc{1}{c|}{$b$}&\mc{1}{c|}{$\Delta \log P$}   \\
        \hline
        \hline
 & RRab+RRc & $-3.11\pm0.26$ & $-0.123\pm0.022$ & $0.197 \pm 0.015$  &$-3.11\pm 0.26$&$-0.122\pm 0.022$ & $0.198 \pm 0.015$\\
        \cline{2-8}
        $J$ & RRab & $-3.45\pm0.39$ & $-0.120\pm0.027$ & - &$-3.46 \pm0.39$&$-0.120\pm 0.027$ & -\\
        \cline{2-8}
        & RRc & $-2.48 \pm 0.47$ & $-0.089\pm0.045$ & - &$-2.47 \pm0.47$&$-0.090 \pm0.046$  & -\\
        \hline
        \hline
        & RRab+RRc & $-3.17\pm0.26$ & $-0.327\pm0.023$ & $0.180 \pm 0.015$ &$-3.18 \pm0.26$&$-0.327 \pm0.023$ & $0.181 \pm 0.015$\\
        \cline{2-8}
        $H$ & RRab & $-3.40\pm0.35$ & $-0.325\pm0.024$ & - &$-3.41 \pm0.35$& $-0.325\pm 0.024$& -\\
        \cline{2-8}
        & RRc & $-2.78\pm0.47$ & $-0.249\pm0.045$ & - &$-2.76 \pm0.47$&$-0.250\pm 0.046$& -\\
        \hline
        \hline
        & RRab+RRc &$-3.38 \pm0.26$&$-0.385\pm0.022$ & $0.178 \pm 0.014$ & $-3.39\pm 0.26$&$-0.385 \pm 0.022$ & $0.179 \pm 0.014$\\
        \cline{2-8}
        $K_s$ & RRab &$-3.59 \pm0.31$&$-0.384\pm 0.022$& - &$-3.61 \pm0.31$&$-0.384 \pm 0.022$ & -\\
        \cline{2-8}
        & RRc &$-3.00\pm 0.46$&$-0.297\pm 0.044$& - &$-2.99\pm 0.47$&$-0.299 \pm0.045$& - \\
        \hline
        \hline
        & RRab+RRc &$-3.57\pm 0.26$&$-0.567\pm 0.022$  &$0.167 \pm 0.014$ &$-3.55\pm 0.26 $&$-0.570\pm 0.022$ & $0.165 \pm 0.014$\\
        \cline{2-8}
        $W_{JK}$ & RRab &$-3.69\pm 0.29$&$-0.566 \pm0.020$& - &$-3.71\pm 0.29$&$-0.565\pm 0.020$ & - \\
        \cline{2-8}
        & RRc &$-3.36 \pm0.46$&$-0.441\pm 0.045$& - &$-3.34 \pm0.48$&$-0.435 \pm0.045$ & -\\
        \hline
        \hline  
        \hline
        \hline

        \mc{2}{|l||}{PL relations} &\mc{3}{c||}{photo-geometric distance}&\mc{3}{c|}{Astrometry-Based Luminosity}\\
        \hline
        \mc{1}{|l|}{band}&\mc{1}{l||}{population}& \mc{1}{c|}{$a$} & \mc{1}{c||}{$b$} &\mc{1}{c||}{$\Delta \log P$} &  \mc{1}{c|}{$a$} & \mc{1}{c|}{$b$}&\mc{1}{c|}{$\Delta \log P$}   \\
        \hline
        \hline
        & RRab+RRc &$-3.09\pm 0.25$&$-0.126\pm 0.022$ & $0.194 \pm 0.015$ &$-3.13 \pm0.24$&$-0.114\pm 0.021$ & $0.204 \pm 0.013$\\
        \cline{2-8}
        $J$ & RRab &$-3.43\pm 0.38$&$-0.124 \pm0.027$& - & $-3.64 \pm0.42$&$-0.117 \pm0.027$ & -\\
        \cline{2-8}
        & RRc &$-2.45\pm 0.46$&$-0.083\pm 0.044$& - & $-2.49 \pm0.45$&$-0.087\pm 0.049$ & -\\
        \hline
        \hline
        & RRab+RRc &$-3.15\pm 0.26$&$-0.330 \pm0.023$& $0.177 \pm 0.015$ &$-3.20\pm 0.27$&$-0.321\pm 0.024$ & $0.184 \pm 0.015$\\
        \cline{2-8}
        $H$ & RRab &$-3.38\pm 0.34$&$-0.328 \pm0.024$& - &$-3.48\pm 0.37$&$-0.322\pm 0.024$ & -\\
        \cline{2-8}
        & RRc &$-2.75\pm 0.47$&$-0.243 \pm0.045$& - &$-2.88\pm 0.45$&$-0.252 \pm0.049$ & -\\
        \hline
        \hline
        & RRab+RRc &$-3.36 \pm0.25$&$-0.388 \pm0.022$& $0.176 \pm 0.014 $ &$-3.41 \pm0.27$&$-0.380\pm 0.024$ & $0.183 \pm 0.014$\\
        \cline{2-8}
        $K$ & RRab &$-3.58\pm 0.30$&$-0.387\pm 0.021$& - &$-3.72\pm 0.33$&$-0.382\pm 0.021$ & -\\
        \cline{2-8}
        & RRc &$-2.97\pm 0.46$&$-0.291 \pm0.044$& - &$-3.05 \pm0.44$&$-0.298\pm 0.048$ & -\\
        \hline
        \hline
        & RRab+RRc &$-3.55\pm 0.25$&$-0.570 \pm0.022$& $0.165 \pm 0.014$ &$-3.62 \pm0.29$&$-0.563\pm 0.026$ & $0.170 \pm 0.015$\\
        \cline{2-8}
        $W_{JK}$ & RRab &$-3.68\pm0.28$&$-0.569\pm 0.019$& - &$-3.78 \pm0.31$&$-0.564\pm 0.020$ & - \\
        \cline{2-8}
        & RRc &$-3.33\pm 0.47$&$-0.435 \pm0.045$& - &$-3.44 \pm0.44$&$-0.443\pm 0.049$ & -\\
        \hline
    \end{tabular}
    \caption{Fitted parameters of PL relations ($a$ - slope, $b$ - intercept, $\Delta \log P$ - shift of $\log P$ of RRc stars) for $JHK_s$ bands and the $W_{JK}$ Wesenheit index using four different techniques mentioned in the text. Relations for RRab+RRc and RRab are fitted using the pivot logarithm of period $log P_0=-0.25$ while relations for RRc are based on $log P_0=-0.45$ in order to minimize intercept errors. \label{tab:PLR}}
\end{table*}

\begin{figure*}
    \centering
    \includegraphics[width=\textwidth]{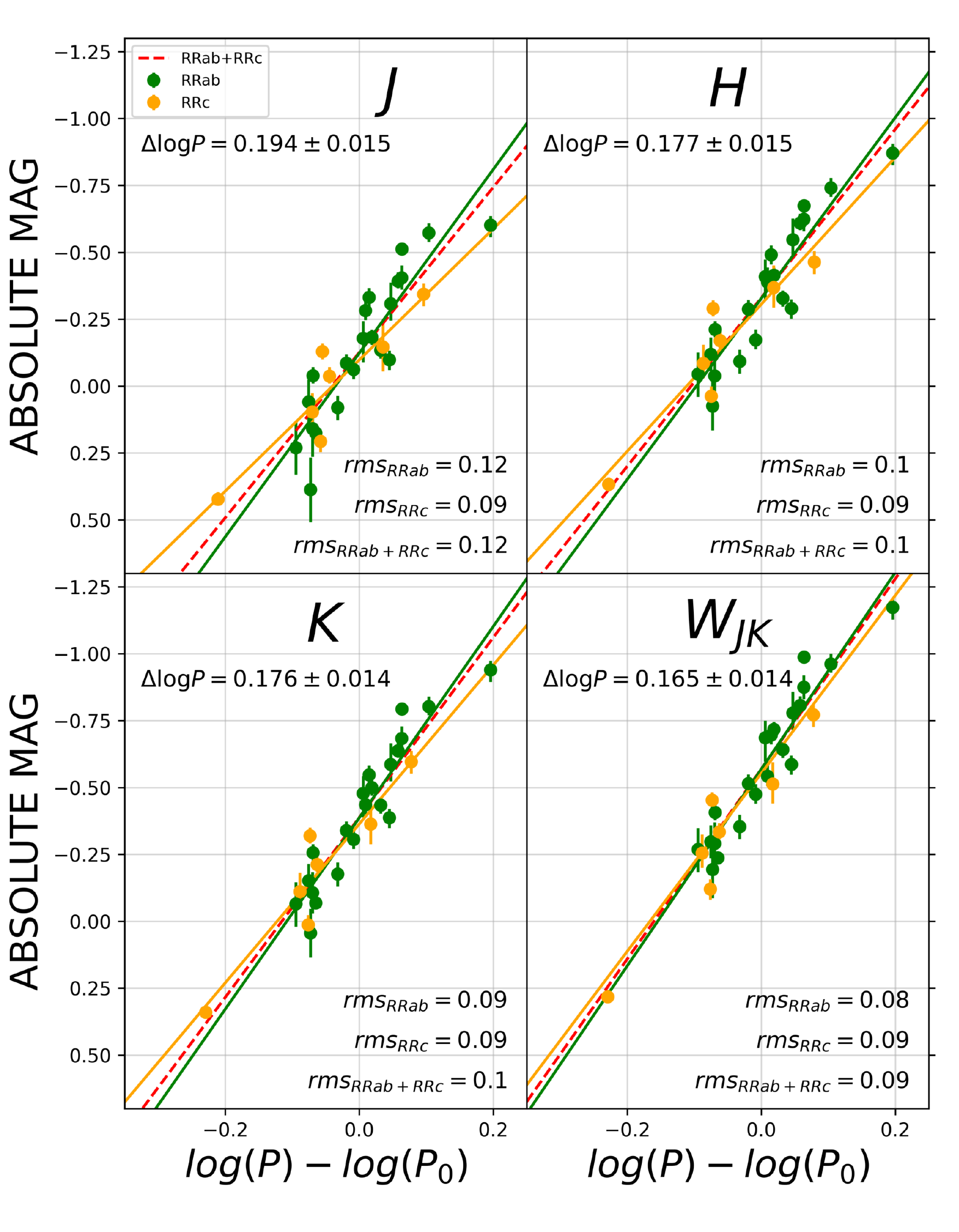}
    \caption{PL relations in $JHK_s$ and the $W_{JK}$ index based on photo-geometric distances of \cite{B-J2021}. Non-symmetric error bars that give upper and lower 1$\sigma$ distances of absolute magnitudes stem from the propagation of the parallax error (dominant) and the photometric error.}
    \label{fig:P-L-BJPG}
\end{figure*}

Table \ref{tab:PLR} includes slopes and intercepts of PL relations established using the four methods and Figure \ref{fig:P-L-BJPG} depicts relations fitted based on the photo-geometric distances of \cite{B-J2021}. While all results obtained from linear fits are practically the same, the ABL usually yields expected parameter values that deviate slightly from the linear cases (but not significantly in the statistical sense, given the uncertainties of fitted parameters). Twenty eight stars were used to establish PL
relations in the $JK_s$ bands and in the $W_{JK}$. In the case of the $H$-band, 27
stars were used (where FW Lup was excluded as it was too bright to perform reliable photometry given that the IRIS camera is the most sensitive in this band). Fits are divided into three subsamples: RRab and RRc type stars, and a sample corresponding to the mixed population (RRab+RRc). We see that values of $\Delta \log P$ obtained independently for different bands are in agreement given their uncertainties.

Error bars corresponding to points in Figure~\ref{fig:P-L-BJPG} are associated with parallax and statistical photometric errors only (where the parallax error is the dominant component). We adopted uncertainties of the fitted parameters obtained in the fitting process. The zero point uncertainty of the IRIS photometry ($0.002$\,mag) is negligible compared to the uncertainties of the intercepts of the PL relations ($\ge 0.02$\,mag).

Although all methods of the establishment of PL relations give consistent, almost identical results with very similar uncertainties (especially for the three approaches based on linear least-squares fits), it is the photo-geometric distance of \citeauthor{B-J2021} and the linear least-squares fit that formally yield the lowest uncertainty values of the fitted parameters.

\subsection{Period-luminosity-metallicity relations}

\begin{table*}[]
    \centering
    \begin{tabular}{|c|c|c|c|c|}
    \hline
    ID & type & $\varpi$ & $r$ (photo-geo) & $\left[\rm Fe/\rm H\right]$\\
    &&[mas]&[pc] & [dex] \\
    \hline
    AE Boo & RRc &$ 1.143 \pm 0.019 $ & $ 874 ^{+ 14 }_{- 16 } $ & $ -1.62 \pm 0.08 $ \\
    AN Ser & RRab &$ 1.026 \pm 0.022 $ & $ 976 ^{+ 19 }_{- 20 } $ & $ 0.05 \pm 0.10 $\\
    BB Eri & RRab &$ 0.722 \pm 0.024 $ & $ 1370 ^{+ 40 }_{- 56 } $ & $ -1.66 \pm 0.040 $\\
    DX Del & RRab &$ 1.758 \pm 0.015 $ & $ 567 ^{+ 5 }_{- 4 } $ & $ -0.19 \pm 0.050 $\\
    EV Psc & RRc &$ 1.133 \pm 0.03 $ & $ 883 ^{+ 28 }_{- 22 } $ & $ - $\\
    FW Lup & RRab &$ 2.800 \pm 0.017 $ & $ 357 ^{+ 2 }_{- 2 } $ & $ -0.17 \pm 0.02 $\\
    HY Com & RRc &$ 0.990 \pm 0.019 $ & $ 1006 ^{+ 18 }_{- 20 } $ & $ -1.75 \pm 0.02 $\\
    IK Hya & RRab &$ 1.299 \pm 0.023 $ & $ 774 ^{+ 16 }_{- 15 } $ & $ -2.54 \pm 0.08 $\\
    MT Tel & RRc &$ 2.070 \pm 0.030 $ & $ 482 ^{+ 6 }_{- 6 } $ & $ -2.58 \pm 0.14 $\\
    RR Leo & RRab &$ 1.084 \pm 0.025 $ & $ 920 ^{+ 23 }_{- 25 } $ & $ -1.58 \pm 0.08 $\\
    RU Psc & RRc &$ 1.278 \pm 0.029 $ & $ 779 ^{+ 20 }_{- 16 } $ & $  - $\\
    RV Cet & RRab &$ 0.976 \pm 0.018 $ & $ 1023 ^{+ 15 }_{- 17 } $ & $ -1.5 \pm 0.02 $\\
    RX Eri & RRab &$ 1.723 \pm 0.023 $ & $ 585 ^{+ 7 }_{- 7 } $ & $ -1.45 \pm 0.15 $\\
    RY Col & RRab &$ 0.993 \pm 0.016 $ & $ 1005 ^{+ 14 }_{- 15 } $ & $ -1.21 \pm 0.02 $\\
    SS Leo & RRab &$ 0.795 \pm 0.025 $ & $ 1261 ^{+ 45 }_{- 36 } $ & $ -1.8 \pm 0.10 $\\
    SV Eri & RRab &$ 1.361 \pm 0.024 $ & $ 733 ^{+ 12 }_{- 10 } $ & $ -2.22 \pm 0.02 $\\
    SX For & RRab &$ 0.868 \pm 0.015 $ & $ 1141 ^{+ 14 }_{- 16 } $ & $ -2.2 \pm 0.02 $\\
    T Sex & RRc &$ 1.340 \pm 0.023 $ & $ 740 ^{+ 11 }_{- 9 } $ & $ -1.52 \pm 0.03 $\\
    U Lep & RRab &$ 0.989 \pm 0.017 $ & $ 1014 ^{+ 16 }_{- 16 } $ & $ -1.81 \pm 0.17 $\\
    UU Vir & RRab &$ 1.281 \pm 0.047 $ & $ 786 ^{+ 32 }_{- 33 } $ & $ - $\\
    V467 Cen & RRab &$ 1.255 \pm 0.023 $ & $ 808 ^{+ 14 }_{- 13 } $ & $  - $\\
    V675 Sgr & RRab &$ 1.199 \pm 0.019 $ & $ 829 ^{+ 13 }_{- 9 } $ & $ -2.47 \pm 0.02 $\\
    V753 Cen & RRc &$ 1.436 \pm 0.014 $ & $ 696 ^{+ 8 }_{- 5 } $ & $ -0.56 \pm 0.04 $\\
    V Ind & RRab &$ 1.506 \pm 0.019 $ & $ 666 ^{+ 9 }_{- 8 } $ & $ -1.62 \pm 0.01 $\\
    WY Ant & RRab &$ 0.979 \pm 0.021 $ & $ 1032 ^{+ 25 }_{- 16 } $ & $ -1.6 \pm 0.15 $\\
    WZ Hya & RRab &$ 1.029 \pm 0.016 $ & $ 974 ^{+ 15 }_{- 15 } $ & $ -1.48 \pm 0.02 $\\
    X Ari & RRab &$ 1.869 \pm 0.019 $ & $ 534 ^{+ 5 }_{- 5 } $ & $ -2.53 \pm 0.08 $\\
    XZ Gru & RRab &$ 0.870 \pm 0.018 $ & $ 1150 ^{+ 17 }_{- 23 } $ & $ - $\\
    \hline
    median  & & $1.28$ & $874$ &$-1.62$ \\
    \hline
    \end{tabular}
    \caption{Parallaxes $\varpi$ from Gaia EDR3 \citep{GAIAEDR3} for Galactic RR~Lyrae stars from our sample corrected with the \cite{LINDEGREN} corrections. Photo-geometric distances $r$ were taken from \cite{B-J2021}, metallicities $\left[\rm Fe/\rm H\right]$ come from \cite{CRESTANI}.}
    \label{tab:metal}
\end{table*}

Twenty three stars from our sample have metallicities reported in \cite{CRESTANI}. Table~\ref{tab:metal} contains the Gaia EDR3 parallaxes, photo-geometric distances \citep{B-J2021}, and metallicities. $H-$ band photometry for FW Lup was once again rejected and 22 stars were used to establish PLZ relation in this case.

\newcommand{\met}{\left[Fe/H\right]}
As our sample of stars is too small to perform 4-parameter fits in the case of PLZ relations, we relied on $\Delta \log P$ values obtained using our PL fits.
We performed three-dimensional fits for PLZ relations based on the photo-geometric distances from \cite{B-J2021}. Unweighted fits were performed using the \textit{curve\_fit} procedure of \cite{Scipy} based on the Levenberg-Marquardt least-squares algorithm and the corresponding parameter uncertainties were estimated by bootstrapping residuals (see Figure \ref{fig:cov}). Assuming the additional dependence of the absolute magnitudes of RRL stars on metallicity:
\begin{equation}
\begin{split}
     M_{\lambda}\left(P', \met' \right)=a_{\lambda}\log P' +b_{\lambda} +\\ + c_{\lambda} \met'
\end{split}
\end{equation}

where $P'$ is based on the same pivot period as before, $\met'=\left(\met - \met_0\right)$, $\met_0=-1.5$ dex is the pivot metallicity that is close to the median metallicity of the sample and $a_{\lambda}$, $b_{\lambda}$, $c_{\lambda}$ are fitted period slope, intercept, and metallicity slope of a relation, respectively. 

\begin{table*}[]
\centering
    \begin{tabular}{|c|c||c|c|c|c|}
        \hline
        \mc{2}{|l||}{PLZ relations} & \mc{4}{|c|}{photo-geometric distance}\\
        \hline
        \mc{1}{|l|}{band}& population & \mc{1}{|c|}{$a$} & \mc{1}{c|}{$b$} & \mc{1}{c|}{$c$} & \mc{1}{c|}{rms} \\
        \hline
        \hline
        & RRab+RRc &$-2.67 \pm  0.33$  & $-0.140 \pm  0.020$ & $0.085 \pm  0.030$ & $0.092$ \\
        \cline{2-6}
        $J$ & RRab & $-3.09 \pm  0.52$  & $-0.147 \pm  0.020$ & $0.070 \pm  0.038$ & $0.087$ \\
        \cline{2-6}
        \hline
        \hline
        & RRab+RRc & $-2.74 \pm  0.28  $  & $-0.331\pm  0.019  $ & $ 0.093 \pm   0.031  $ & $0.082$ \\
        \cline{2-6}
        $H$ & RRab & $-2.84 \pm  0.44  $  & $-0.343 \pm  0.019  $  & $0.084 \pm  0.036  $ & $0.076$ \\
        \cline{2-6}
        \hline
        \hline
        & RRab+RRc & $-3.03 \pm  0.25  $ & $-0.396 \pm  0.016  $ & $0.083\pm  0.025  $ & $0.073$ \\
        \cline{2-6}
        $K$ & RRab & $-3.10 \pm  0.39  $ & $-0.404 \pm  0.016  $ & $0.080 \pm  0.029  $ & $0.066$ \\
        \cline{2-6}
        \hline
        \hline
        & RRab+RRc &$-3.26 \pm  0.23  $  & $-0.573 \pm  0.015  $ & $0.083 \pm  0.024  $ & $0.070$ \\
        \cline{2-6}
        $W_{JK}$ & RRab & $-3.11 \pm  0.37  $  & $-0.581 \pm  0.015  $ & $0.087 \pm  0.028  $ & $0.064$ \\
        \hline
    \end{tabular}\par
    \caption{Fitted parameters of PLZ relations ($a$ - slope, $b$ - intercept, $c$ - metallicity slope) for $JHK_s$ bands and the $W_{JK}$ Wesenheit index based on the photo-geometric distances of RR Lyrae stars published by \cite{B-J2021}. All relations are fitted using a pivot logarithm of period $log P_0=-0.25$ and a pivot metallicity of $\met_0=-1.5$ dex. \label{tab:PLZR}}
\end{table*}

\begin{figure*}
    \centering
    \includegraphics[width=0.8\textwidth]{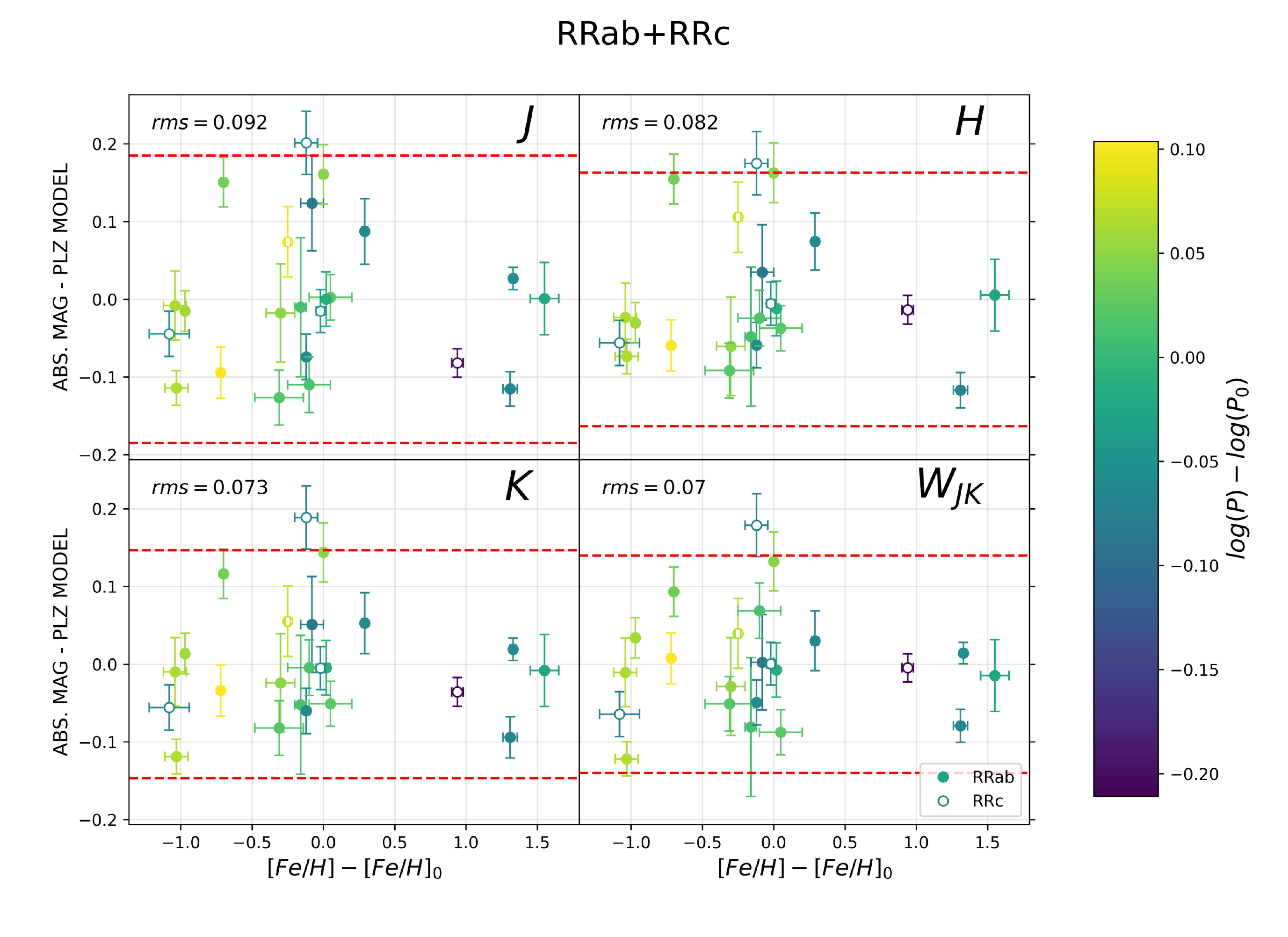}
    \caption{Residuals of plane fits for three bands and the Wesenheit index for the mixed population (RRab+RRc) and separately for the population of fundamental pulsators (RRab). Dashed lines denote the $2\times$rms deviation from the model; $\log P_0=-0.25$ and $\met_0=-1.5$.}
    \label{fig:Res-RRab+RRc}
\end{figure*}
\begin{figure*}
    \ContinuedFloat
    \centering
    \includegraphics[width=\textwidth]{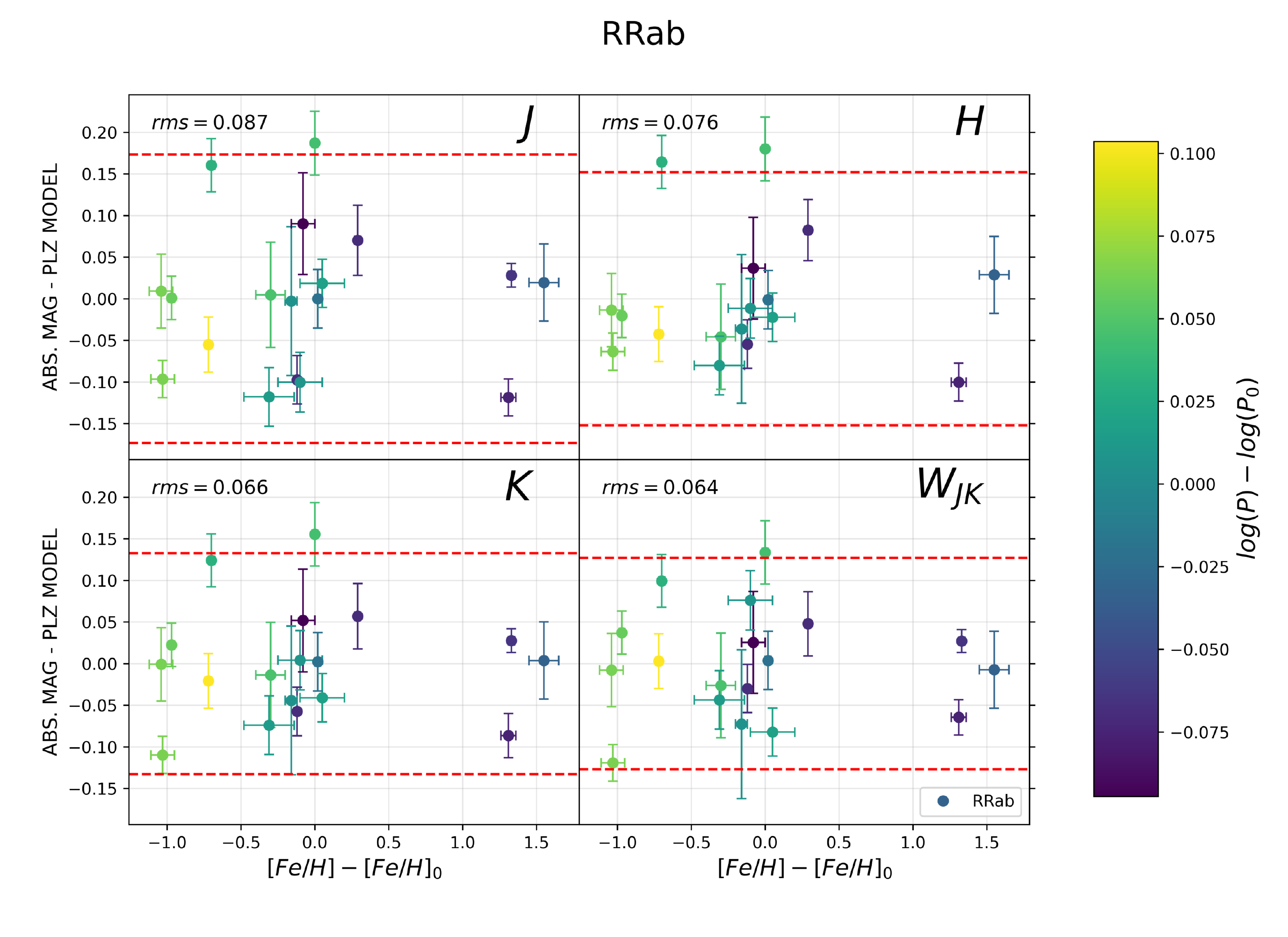}
    \caption{continued - fundamental pulsators only}
    \label{fig:Res-RRab}
\end{figure*}

Table \ref{tab:PLZR} presents fitted parameters of the PLZ relations for $JHK_s$ bands and the Wesenheit index. Figure \ref{fig:Res-RRab+RRc} depicts the residuals of our fits to the PLZ relations for the mixed population (RRab+RRc) and the population of fundamental pulsators (RRab) separately.

We have also fitted PLZ relations based on the ABL and obtained results that are identical in the statistical sense.

The fitted relations indicate smaller metallicity dependence than in the theoretical works of \cite{Bono2001} ($0.17$\,mag/dex in $K$), \cite{CATELAN-PLZ} ($0.17-0.19$\,mag/dex for $JHK$), \cite{MARCONI} ($0.15-0.19$\,mag/dex for $JK$ and $W_{JK}$, with lower values corresponding to the RRc stars), and \cite{BRAGA} ($0.16-0.19$\,mag/dex for $W_{JK}$).
On the other hand, empirical calibrations usually give smaller dependence on metallicity\footnote{Literature calibrations without distinction between populations of fundamental and first-overtone pulsators are given for the mixed population of RRab+RRc with the applied procedure of the fundamentalization of periods of RRc stars based on \cite{Iben0127} where $\Delta \log P=0.127$.}, in a very good agreement with values reported in this work. \cite{Sollima2006} obtained $(0.08\pm0.11)$\,mag/dex for the $K_s-$band. \cite{MURAVEVA2015} found $(0.03 \pm 0.07)$\,mag/dex for the $K_s-$band \textit{Vista Magellanic Cloud Survey (VMC)} photometry of RR Lyrae stars from the LMC, explaining the especially low parameter value by a narrow span of their metallicities in that galaxy. They reported $(0.07 \pm 0.04)$\,mag/dex for the Galactic RR~Lyrae stars in the same work. \cite{VMC}, who utilized VMC photometry of 22 thousand RR Lyrae stars from the LMC, report effects of $0.096 \pm 0.004$\,mag/dex and $0.114 \pm 0.004$\,mag/dex for the mixed and the RRab populations in $K_s-$ band, respectively.

Figure \ref{fig:cov} depicts covariance between fitted parametrs of the PLZ relation for fundamental pulsators in the $K-$ band obtained through bootstrapping residuals of the original fit where absolute magnitudes were calculated based on the photo-geometric distaces of \cite{B-J2021}. The utilization of the pivot period and metallicity allows for minimization of the correlation between fitted slopes and the intercept. However, we observe a strong correlation between the period slope and the metallicity slope. Such correlation was reported recently by \cite{MULLEN} for a much larger sample of more than 1000 Galactic RR Lyrae stars in the case of the mid-infrared band (their Figure 3).

\begin{figure*}
	\centering
	\includegraphics[width=\textwidth]{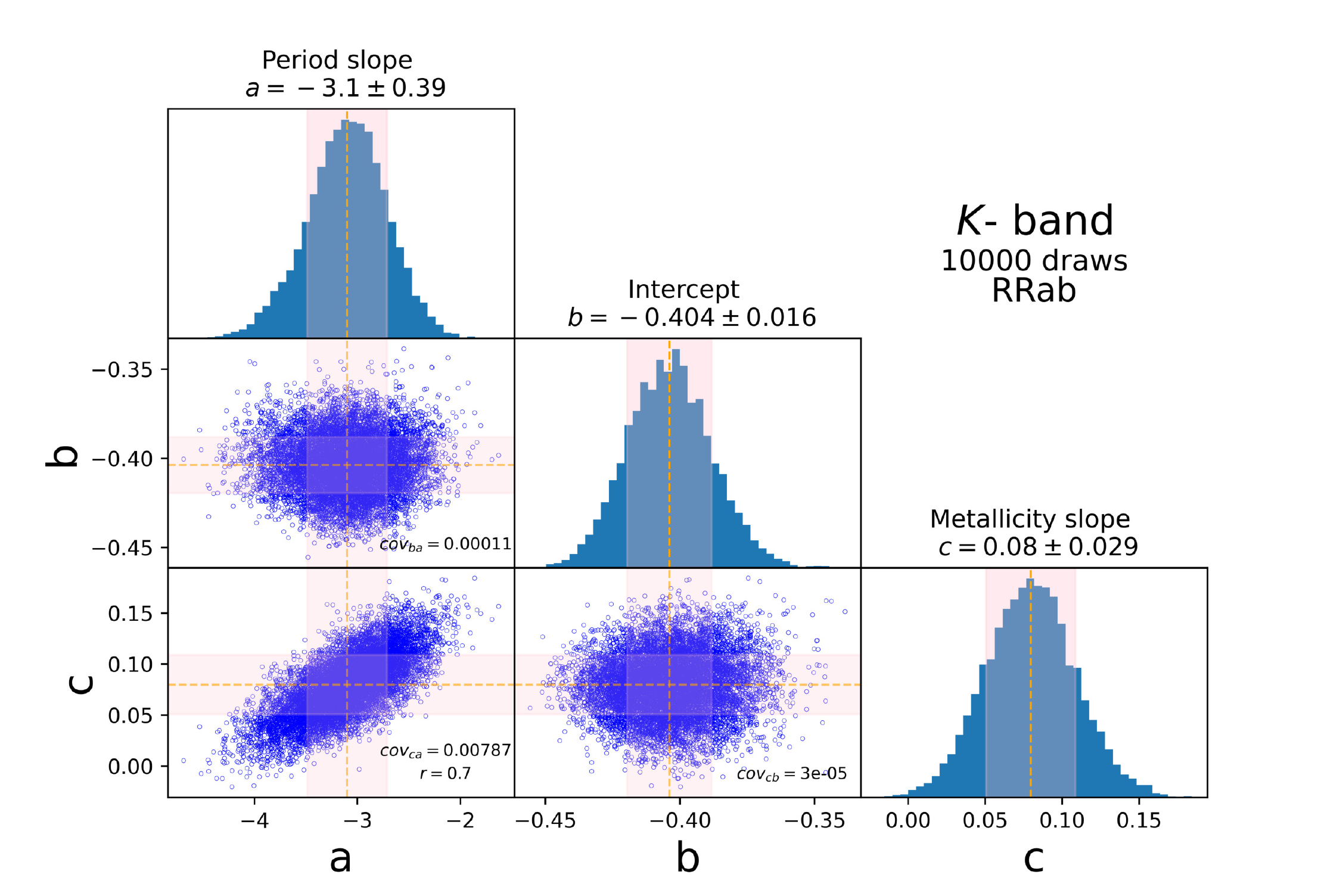}
	\caption{Graphic depiction of covariance between fitted parameters of a PLZ relation for the exemplary case of the $K-$ band and the RRab population obtained by bootstrapping residuals.}
	\label{fig:cov}
\end{figure*}

We note here that the spread of residuals of PLZ relations is relatively low compared to, e.g., relations presented by \cite{NEELEY} (rms of around 0.18 mag). Our relations are based on Gaia DR3 parallaxes that have improved precision and accuracy compared to previous data releases.

\section{Discussion}

\subsection{The comparison of zero points of PLZ relations with other calibrations.}

When it comes to accurate distance determinations using PL or PLZ relations for RR~Lyrae stars (or pulsating stars in general), the most important challenge is to accurately define the zero point of a calibration. It is especially challenging to estimate the portion of the zero point uncertainty that is associated with elusive systematic errors. This is why we should compare our results with other calibrations, especially those obtained using independent, alternative methods.

We compare our results with calibrations derived in the last years (\citealt{MURAVEVA2015}, \citealt{MURAVEVA_GDR2}, \citealt{NEELEY} \citealt{VMC}). However, we begin with the comparison of our calibration with \cite{SOLLIMA}, which was extensively used as a fiducial phenomenological relation used for the purpose of distance determinations to nearby galaxies by \cite{SCULPTOR}, \cite{SZEWCZYK-LMC}, \cite{SZEWCZYK-SMC}, \cite{CARINA}, \cite{FORNAX}.

The relation reported by \cite{SOLLIMA} ($K_s-$ band of the 2MASS system):
\begin{equation}
\begin{split}
    M_{K_s}=(-2.38 \pm 0.04)\log P + \\ + (0.08 \pm 0.11) \met - (1.07 \pm 0.11)
\end{split}
\end{equation}
has its zero point established based on the HST trigonometric parallax \citep{RRLPAR} of a single star: the prototype RR Lyr. After fixing both our period and the metallicity slope to the values from that work, we fit a zero point to our data and obtain the value of $(-0.862 \pm 0.019)$\,mag  (statistical uncertainty). The two values are in agreement, although only given the substantial total error reported by \cite{SOLLIMA}. This is also due to the difference between the parallaxes of RR Lyr from HST reported by \cite{RRLPAR} ($\varpi_{HST}=3.82 \pm 0.02$ mas) and from Gaia EDR3 ($\varpi_{GEDR3}=4.00 \pm 0.03$ mas, including the $-0.02$ mas correction of \citealt{LINDEGREN}). The $\varpi_{GEDR3}-\varpi_{HST}$ difference corresponds to an absolute magnitude and zero point shift of $0.1$ mag.

\cite{MURAVEVA2015} established RRL PLZ relations for the LMC in the VISTA $K_s$--band\footnote{Our $K_s-$band relations are calibrated onto the 2MASS $K_s$--band, which is shifted by just about $3-4$\,mmag from VISTA $K_s$ for the typical color of RR~Lyrae stars \citep{MURAVEVA2015}. For the formal agreement, we are still converting the photometry to the VISTA system using the transformation equations given at \url{http://casu.ast.cam.ac.uk/surveys-projects/vista/technical/photometric-properties} \citep{Gonzalez-Fernandez}.} using VMC photometry of 70 stars from \cite{CIONI}. They tied their calibrations to two different anchors. The first one being the very accurate $2\%$ distance to the LMC  by \cite{LMC-2pDEB}. In an alternative approach, they assumed the same period and metallicity slopes as in the LMC and performed a one-parameter fit to 4 Galactic RR~Lyrae stars having trigonometric parallaxes determined using the HST by \cite{BENEDICT}. Those two zero points do not agree with each other, which may be caused by the difficulties of the complex analysis of relative parallaxes from the HST. The relation presented by \cite{MURAVEVA2015} based on the anchoring distance of \cite{LMC-2pDEB} is as follows:

\begin{equation}
\begin{split}
    M_{K_s}=(-2.73 \pm 0.25) \log P + \\ + (0.03 \pm 0.07) \met  - (1.06 \pm0.01)
\end{split}
\end{equation}

while the zero point based on the parallaxes of 4 stars by \cite{BENEDICT} is $(-1.25 \pm 0.06)$\,mag instead.

At first, we notice that both the period and the metallicity slope are in agreement with those derived in the present study for the $K_s-$band and the mixed population. In order to compare zero points of our calibrations, we fix both slopes and perform a fit of one parameter once again. The ABL fit yields $(-1.015 \pm 0.020)$\,mag while the fit based on the photo-geometric distances gives $(-1.034 \pm 0.020)$\,mag. Besides the agreement of these zero points with the LMC-anchored calibration of \citeauthor{MURAVEVA2015}, the two types of fit give very similar results when it comes to the determination of the zero points of the PLZ relations.

\cite{MURAVEVA2015} also presented another calibration based on 23 Galactic RR~Lyrae stars that were used for studies associated with the Baade-Wesselink (B-W) method (\citealt{JONES1988}, \citeyear{JONES1992}, \citealt{FERNLEY1990}, \citealt{LIU}, \citealt{CACCIARI}, \citealt{SKILLEN}, \citealt{FERNLEY1994}). Using photometry and reddening from \cite{FERNLEY1998}, the authors estimated stellar absolute magnitudes using a fixed value of the projection factor $p=1.38$. The assumed $p$ value makes the zero point of these absolute magnitudes arbitrary, as no robust calibration of $p-$ factors for RR~Lyrae stars was performed. Even though the relation is calibrated in the Johnson photometric system, authors underline the average difference with the 2MASS $K_s$--band of the order of $0.03$\,mag while the B-W-based absolute magnitudes have uncertainties of $0.15-0.25$\,mag. Finally, \cite{MURAVEVA2015} obtained a Galactic relation in the following form:

\begin{equation}
\begin{split}
    M_{K_s}=(-2.53 \pm 0.36) \log P + \\ +(0.07 \pm 0.04) [Fe/H]-(0.95 \pm 0.14)
\end{split}
\end{equation}

The metallicity dependence is slightly larger now and in better agreement with the value derived in this work. However, the zero point uncertainty is much larger compared to the LMC calibration. The fit of zero point to our data after fixing the above slopes yields $-0.918 \pm 0.018$\,mag and is in good agreement despite the relatively low precision of the Galactic calibration based on the B-W distances. 

In later work, \cite{MURAVEVA_GDR2} included the zero point of the PLZ relation based on the Gaia DR2 parallaxes. They used Bayesian modeling, with the parallax systematic offset being the parameter of  the model. The authors found a systematic offset of the Gaia DR2 parallaxes of $-0.054$\,mas in the case of the PLZ relation for the $K_s-$band derived from a sample of 400 stars from the Milky Way. All these constraints resulted in a distance modulus of the LMC (using the same sample of RR~Lyrae stars from the LMC as in \citealt{MURAVEVA2015}) $\mu_{\rm LMC} = 18.55 \pm 0.11$\,mag. Their relation (2MASS system) takes the following form:

\begin{equation}
\begin{split}
    M_{K_s}=(-2.58 \pm 0.20) \log P + \\ + (0.17 \pm 0.03) [Fe/H] - (0.84 \pm 0.09)
\end{split}
\end{equation}

The metallicity dependence is even larger in that analysis but still in agreement with the value obtained in this work. After fixing period and metallicity coefficients to their literature values, we obtain a zero point value of $-0.775 \pm  0.019$ which is consistent with the zero point of \cite{MURAVEVA_GDR2}.

\cite{NEELEY} presents PL and PLZ relations for RR~Lyrae stars based on Gaia DR2 parallaxes. They used the photometry of 55 stars from our Galaxy gathered for the \textit{Carnegie RR~Lyrae Program}. The authors obtained a rather large scatter of residuals of their fits ($\sim 0.2$\,mag), which they identified as possibly due to unaccounted uncertainties or systematics. The work includes a variety of different PL, PLZ, PW, and PWZ relations (including Wesenheit indices instead of luminosity in a given band) obtained using weighted least-squares fits. The authors also use a Bayesian approach and a robust analysis (including the weighting of points based on the scatter of the fit) but find no significant differences between different methods.

The results (2MASS photometric system) of \cite{NEELEY} are as follows:

\begin{equation}
\begin{split}
    M_J=(-1.91 \pm0.29)(\log P +0.3) + \\ +(0.20\pm0.03)([Fe/H]+1.36)-(0.14 \pm 0.02)
\end{split}
\end{equation}
\begin{equation}
\begin{split}
    M_H=(-2.40 \pm0.29)(\log P +0.3) + \\ + (0.17\pm0.03)([Fe/H]+1.36)-(0.31 \pm 0.02)
\end{split}
\end{equation}
\begin{equation}
\begin{split}
    M_{K_s}=(-2.45 \pm0.28)(\log P +0.3) + \\ + (0.17\pm0.03)([Fe/H]+1.36)-(0.37 \pm 0.02)
\end{split}
\end{equation}
\begin{equation}
\begin{split}
    W_{JK}=(-2.91 \pm0.30)(\log P +0.3) + \\ + (0.15\pm0.03)([Fe/H]+1.36)-(0.53 \pm 0.02)
\end{split}
\end{equation}

Period and metallicity slopes are systematically larger compared to our calibrations but still consistent within the uncertainties. As \citeauthor{NEELEY} show in their Figure~9, all fitted parameters, slopes, and the zero point (intercept) of PLZ relations depend monotonically on the zero point offset of Gaia DR2 parallaxes, where this value may range from $-0.030$\,mas to $-0.056$\,mas and may be investigated using quasars \citep{OFFSET_GDR2}\footnote{\cite{OFFSET_GDR2} report an offset of $(-0.056 \pm 0.005)$\,mas for RR~Lyrae stars -- a value which is consistent with the result of \cite{MURAVEVA_GDR2}. \cite{NEELEY} note that it is not possible to set such an offset for RR~Lyrae stars without assuming a PL relation.}. The authors adopt a smaller offset (in terms of the absolute value) than \cite{MURAVEVA_GDR2}, i.e., $-0.03$\,mas, which is the same as the offset from \cite{BAILER2018}.

Using the same pivot period and metallicity ($\log P_0=-0.3$ and $[Fe/H]_0=-1.36$) as \cite{NEELEY} and fixing our slopes to the values obtained in that work, we get zero point values of $-0.001 \pm 0.024$, $-0.173 \pm 0.020$, $-0.235 \pm 0.019$, and $-0.396 \pm 0.017$ for $J$, $H$, $K$, and $W_{JK}$, respectively. The zero points of our relations are systematically larger by about $0.14$\,mag. Interestingly, the authors report that calibrations based on the adoption of a parallax offset of $-0.06$\,mas yield an LMC distance that is $0.1$\,mag smaller than without an offset. However, we keep in mind that \cite{NEELEY} used Gaia DR2 parallaxes and our research is based on EDR3. The parallax zero points (and their corrections) for different data releases are different: the parallax zero points published in the EDR3 are improved compared to DR2.

Another calibration takes advantage of the VMC photometry of RR~Lyrae stars from the LMC. \cite{VMC} established PL and PLZ relations based on 22 thousand stars -- among them almost 17,000 fundamental pulsators. The measurements were taken in, i.a., $J$ and $K_s$ and comprised a mixed population.  We fitted PL relations for our sample of RRLs with slopes adopted from \cite{VMC}. Intersections were fitted separately for RRab, RRc, and RRab+RRc samples. Intersections obtained using fits with fixed slopes serve for distance determinations. Namely, the LMC distance modulus is the difference between the intercept value reported in \cite{VMC} and the intercept obtained through our fit. In the case of PLZ relations, we have additionally fixed metallicity slopes to the values reported by \cite{VMC}.

Results, summarized in Tables \ref{tab:PL_CUSANO} \& \ref{tab:PLZ_CUSANO} are in very good agreement with the canonical distance from eclipsing binaries of $\mu_{\rm LMC}=18.477 \pm 0.004 \pm 0.026$\,mag of \cite{LMC-DEB}. They also agree with the median distance of RR Lyrae stars from the LMC of \cite{JACYSZYN}: $\mu_{LMC}=18.522 \pm 0.063$ mag based on the $W_{VI}$ Wesenheit index and a theoretical PWZ relation from \cite{BRAGA}. In these tables, $\delta_{\rm stat}$ is a statistical error of the fit of the zero point of the relation while $\delta_{\rm LMC}$ is a superposition of errors associated with intercept fits, $\delta_{\rm LMC}=\sqrt{\delta_{\rm stat}^2+\delta_{\rm VMC}^2}$, where $\delta_{\rm VMC}$ is the uncertainty of the zero point of the fiducial relation reported by \cite{VMC}. $\delta_{\rm VMC}$ is one of the systematic errors that is involved in the determination. Obviously, it is usually negligible compared to $\delta_{\rm stat}$; however, this is not the case for RRc stars. Neither the use of different populations of RR~Lyrae stars nor the inclusion of the metallicity dependence changes the value of the distance in this case. All determinations, based both on PL and PLZ relations, are in agreement given the small statistical uncertainties of the zero point fits. However, as expected, uncertainties and residuals are usually smaller in the case of PLZ relations.

\begin{table*}
    \centering
    \begin{tabular}{|c|c||c|c|c|c|c|c|}
    \hline
   band & population & slope & $\mu_{\rm LMC}$ & $\delta_{\rm stat}$ & $\delta_{\rm VMC}$ & $\delta_{\rm LMC}$ & rms \\
        &       &     &  [mag] & [mag]    & [mag] & [mag] &[mag] \\
   \hline
   \hline
                & RRab+RRc & $-2.00$ & 18.420 & 0.028 & 0.004 & 0.028 & 0.14\\
                \cline{2-8}
    $J$         & RRab & $-2.50$ & 18.454 & 0.030 & 0.005 & 0.030 & 0.13\\
                \cline{2-8}
                & RRc & $-2.53$ & 18.465 & 0.037 & 0.022 & 0.043 & 0.09\\
    \hline
    \hline
                & RRab+RRc & $-2.53$ & 18.442 & 0.022 & 0.004 & 0.023 & 0.12\\
                \cline{2-8}
    $K_s$         & RRab & $-2.84$ & 18.462 & 0.024 & 0.005 & 0.024 & 0.11\\
                \cline{2-8}
                &  RRc & $-2.98$ & 18.442 & 0.037  & 0.023 & 0.044 & 0.09\\
    \hline
    \hline
                & RRab+RRc & $-2.888$ & 18.473 & 0.020 & 0.004  & 0.020 & 0.10\\
                \cline{2-8}
    $W_{JK}$    & RRab & $-3.075$ & 18.492 & 0.021 & 0.006 & 0.022 & 0.09\\
                \cline{2-8}
                & RRc & $-2.289$ & 18.514 & 0.038 & 0.055 & 0.067 & 0.09\\
    \hline
    \end{tabular}
    \caption{LMC distance moduli $\mu_{\rm LMC} \pm \delta_{\rm LMC}$ obtained by fitting the zero point of a relation while keeping the period slopes fixed to the values from \cite{VMC}.}
    \label{tab:PL_CUSANO}

    \centering
    \begin{tabular}{|c|c||c|c|c|c|c|c|c|}
    \hline
   band & population & period & metallicity & $\mu_{\rm LMC}$ & $\delta_{\rm stat}$ & $\delta_{\rm VMC}$ & $\delta_{\rm LMC}$ & rms \\
        &            &   slope    & slope & [mag] & [mag] &[mag] & [mag] & [mag] \\
   \hline
   \hline
                & RRab+RRc & $-1.91$ & 0.095 & 18.428 & 0.022 & 0.007 & 0.024 & 0.105\\
                \cline{2-9}
    $J$         & RRab & $-2.45$ & 0.121 & 18.465 & 0.022 &0.008 & 0.024 & 0.091\\
    \hline
    \hline
                & RRab+RRc & $-2.41$ & 0.096 & 18.438 &0.020 & 0.007 & 0.021 & 0.093\\
                \cline{2-9}
    $K_s$         & RRab & $-2.80$ & 0.114 & 18.479 & 0.017 & 0.008 & 0.019 & 0.069\\
    \hline
    \hline
                & RRab+RRc & $-2.810$ & 0.094 & 18.456 &0.020 & 0.008 & 0.022 & 0.095\\
                \cline{2-9}
    $W_{JK}$    & RRab & $-3.033$ & 0.111 & 18.502 & 0.016 & 0.009 &0.018 & 0.065\\
    \hline
    \end{tabular}
    \caption{Same as Table~\ref{tab:PL_CUSANO} but based on PLZ relations with the metallicity slope also set to a fixed value.}
    \label{tab:PLZ_CUSANO}
\end{table*}

Recently, \cite{BPLZ} published PLZ relations for near-infrared bands based on 964 RRL stars from 11 Galactic globular clusters and 346 field Galactic RRL stars. Field stars, having their apparent magnitudes reported in the 2MASS catalog and distances based on Gaia DR3 parallaxes given by \cite{B-J2021}, allowed authors for the calibration of zero points of global relations and determinations of distances to individual globular clusters. The authors used mean metallicities of globular clusters reported by \cite{Carretta} and spectroscopic metallicites of field stars from the same sources as \cite{MURAVEVA_GDR2} for their sample of around 400 stars. The authors report the following near-infrared PLZ relations calibrated onto the 2MASS photometric system:

\begin{equation}
\begin{split}
M_J=(-1.83 \pm 0.02)\log P+ \\ +(0.20 \pm 0.02)[Fe/H]-(0.44 \pm 0.03)
\end{split}
\end{equation}

\begin{equation}
\begin{split}
M_H=(-2.29 \pm 0.02)\log P+ \\ +(0.19 \pm 0.01)[Fe/H]-(0.74 \pm 0.02)
\end{split}
\end{equation}

\begin{equation}
\begin{split}
M_{K_s}=(-2.37 \pm 0.02)\log P+ \\ +(0.18 \pm 0.01)[Fe/H]-(0.80 \pm 0.02)
\end{split}
\end{equation}

\begin{equation}
\begin{split}
W_{JK}=(-2.73 \pm 0.02)\log P+ \\ +(0.16 \pm 0.02)[Fe/H]-(1.06 \pm 0.03)
\end{split}
\end{equation}

Similarly to \cite{NEELEY}, period slopes and metallicity coefficients are systematically larger compared to our calibrations, which again corresponds to the nonvanishing covariance between the slopes as seen in Figure \ref{fig:cov}. As before, in order to compare zero points of calibrations, we perform fits to our data after fixing period and metallicity slopes given by \cite{BPLZ}. The intercepts we obtain are $-0.28 \pm 0.03$ mag, $-0.60 \pm 0.02$ mag, $-0.70 \pm 0.02$, and $-1.00 \pm 0.02$ for $J-$, $H-$, $K-$ bands, and the $W_{JK}$ index, respectively. We may see that the zero point offsets are significant even though they rely on the same source of parallaxes taken with the same corrections of \cite{LINDEGREN}. The zero point differences between our calibrations and those presented by \cite{BPLZ} range from $0.06 \pm 0.04$ mag for the $W_{JK}$ index, through $0.1 \pm 0.03$ mag for the $K_s-$ band and $0.14 \pm 0.03$ mag for the $H-$ band, to $0.16 \pm 0.04$ mag for the $J-$ band.
Indeed, the authors report the LMC distance based on the photometry of \cite{VMC} and their $K_s-$ band calibration of $\mu_{LMC}=18.54 \pm 0.13$ mag which is $0.1$ mag larger (but still in good agreement) than what we obtain based on our and \cite{VMC} calibrations (see the Table \ref{tab:PLZ_CUSANO}).
Contrary to the comparison of our PLZ calibrations with \cite{NEELEY}, the offset value depends significantly on the band in this case The non-uniform offset is likely caused by the different treatment of reddening.

Our calibration is based on Gaia parallaxes for a sample of RRL stars from the direct neighborhood of the Sun while \cite{BPLZ} relies on a much larger sample of those stars that are on average more distant\footnote{The mean and the median distance (calculated as the inverse of GDR2 parallaxes) of RRL stars from the \cite{MURAVEVA_GDR2} sample is 2.50 kpc and 2.01 kpc, respectively. The median distance of stars from our sample is 874 pc (Table \ref{tab:metal})}. It is worth noting that, as shown by \cite{NEELEY} in the case of Gaia DR2 parallaxes (their Figure 2), the inversion of parallaxes matches distances of \cite{BAILER2018} up to around 1.5 kpc where it starts to be discrepant (Bayesian distances become larger).

Table \ref{tab:comparison} includes differences based on the independently fitted shift of $\Delta \log P=0.176$ (Table \ref{tab:PLR}, Figure \ref{fig:P-L-BJPG}) of RRc stars for the $K-$ band presented in this work, as well as $\Delta \log P=0.127$ applied in all quoted works. Naturally, when it comes to the comparison of zero points the agreement is better when applying the same value of the shift as in the literature. Otherwise, we obtain differences that are about $0.03$ mag larger.

\begin{table*}
    \centering
    \begin{tabular}{|c|c|c|c|}
    \hline
         work & anchor & $\Delta b$ $(\Delta \log P=0.176)$ & $\Delta b$ $(\Delta \log P=0.127)$ \\
         \hline
         \hline
         \cite{SOLLIMA} & HST parallax of RR Lyr \footnote{\cite{RRLPAR}} & $0.21 \pm 0.11$ & $0.18 \pm 0.11$\\
         \hline
         \cite{MURAVEVA2015} & HST parallaxes of 4 stars \footnote{\cite{BENEDICT}} & $ 0.22\pm 0.07$& $0.19 \pm 0.07$\\
         \hline
         \cite{MURAVEVA2015} & DEB in the LMC \footnote{\cite{LMC-2pDEB}} & $0.03 \pm 0.03$&$0.00 \pm 0.03 $  \\
         \hline
         \cite{MURAVEVA2015} & B-W dist. (23 Galactic RRL stars) & $0.03 \pm 0.14$ & $ 0.01 \pm 0.14$ \\
         \hline
         \cite{MURAVEVA_GDR2} & Bayesian determination of GDR2 $\Delta \varpi$ & $0.07 \pm 0.09$ &$0.04 \pm 0.09 $\\
         \hline
         \cite{NEELEY} & GDR2 - $\Delta \varpi$ as in BJ2018\footnote{\cite{BAILER2018}}  & $0.14 \pm 0.03$ & $0.11 \pm 0.03$\\
         \hline
         \cite{VMC} & DEB in the LMC \footnote{\cite{LMC-DEB}}& $0.039 \pm 0.032$ & $0.004 \pm 0.032$\\
         \hline
	 \cite{BPLZ} & distances of BJ2021\footnote{\cite{B-J2021}} based on GDR3 & $ 0.10 \pm 0.03 $ & $0.07 \pm 0.03$\\
	 \hline
    \end{tabular}
    \caption{Differences of zero points of PLZ relations ($\Delta b$, in mag) between calibration presented in this work and selected observational calibrations available in the literature ($K_s-$ band, RRab+RRc) assuming slopes given therein. Uncertainties are superpositions of zero point errors reported in the literature, statistical errors of fits (see the text), and the systematic zero point error of our fits ($0.02$ mag).}
    \label{tab:comparison}
\end{table*}

\subsection{The distance to the LMC}
As discussed in the previous subsection when comparing our calibrations with that of \cite{VMC}, determinations of distances to the LMC based on both PL and PLZ relations yield very similar values. While the metallicity coefficient of PLZ relations is a subject of discussions with different authors reporting different values, it is expected that the distance to the LMC obtained using PL and PLZ relations for different bands presented in this work will be consistent. Our pivot metallicity $[Fe/H]_0=-1.5$ dex is very similar to the mean metallicity of the LMC, $[Fe/H]_{LMC}=-1.48$ dex (as used in \citealt{SZEWCZYK-LMC}). In this sense, the difference of $0.1$ mag/dex in the value of the metallicity coefficient of a PLZ relation influences the LMC distance in a negligible way as it yields an error of $2$ mmag.

Here, we took advantage of the Araucaria photometry of RRL stars from the LMC published by \cite{SZEWCZYK-LMC}. The procedure of obtaining a distance modulus is as follows: we assumed slopes and period shifts from our calibrations and fixed them. In the case of PLZ relations, we assumed a single value of $[Fe/H]_{LMC}=-1.48$. The product of the metallicity slope and the said metallicity acts as a correction of the zero point that is marginal in this case. Subsequently, we fitted zero points of our relations to the period-luminosity diagrams for the LMC RRL stars.

Having covariances between parameters corresponding to our calibrations, we were able to estimate the covariance between intercepts of relations corresponding to the calibrating and the LMC sample. This covariance is needed to properly estimate the variance of the distance modulus \citep{FEIGELSON}:

\begin{equation}
\begin{split}
var(\mu)=var(b_{LMC})+var(b_{CAL})- \\ -2cov(b_{LMC},b_{CAL})
\end{split}
\label{eq:covmu}
\end{equation}
where $b_{LMC}$ and $b_{CAL}$ are intercepts of relations corresponding to the LMC and the calibrating sample, respectively. The covariance $cov(b_{LMC},b_{CAL})$ was estimated by drawing parameters of calibrated PL and PLZ relations from three-dimensional distributions using random.multivariate\_normal function of NumPy \citep{NUMPY} given parameters' expected (mean) values and covariances\footnote{see Table \ref{tab:covs} in the Appendix}. In cases of distances based on PL relations for RRab or RRc, draws were performed using two-dimensional distributions (without the need to draw the $\Delta \log P$). 

\begin{figure*}
    \centering
    \includegraphics[width=\textwidth]{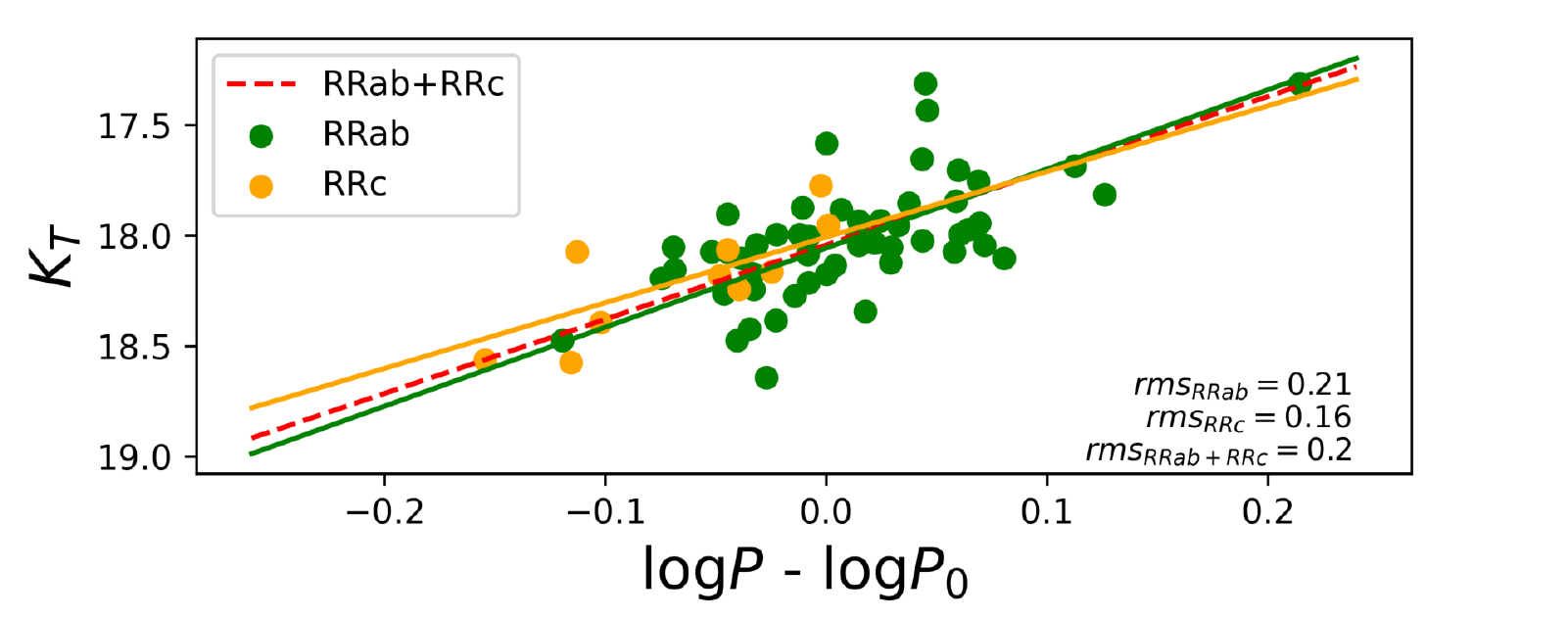}
    \caption{The fit of intercepts of PL relations for the LMC assuming fiducial slopes of PL relations for the $K-$ band presented in this work and mean photometry of RRL stars based on light curve templates taken from \cite{SZEWCZYK-LMC}.}
    \label{fig:PL_LMC}
\end{figure*}

Tables \ref{tab:PL_SZEWCZYK} and \ref{tab:PLZ_SZEWCZYK} present distance moduli of the LMC depending on the photometric band and the mode of pulsations based on the fiducial PL and PLZ relations presented in this work, respectively. \cite{SZEWCZYK-LMC} present mean $J-$ and $K-$ band magnitudes of RRL stars from the LMC and we constructed the corresponding Wesenheit $W_{JK}$ indices based on them. Additionally, they present mean $K-$ band magnitudes based on templates of light curves ($K_T$). We immediately notice that obtained distance moduli based on the mixed and RRab populations are in good agreement for each photometric band and the $W_{JK}$ index. Distances based on RRc stars are systematically smaller, but their uncertainties are larger due to the sizes of both the calibrating and the LMC sample of those stars. Obtained distance moduli are also in good agreement with distances based on the VMC photometry of \cite{VMC} and they agree well with the distance to the LMC based on detached eclipsing binaries of \cite{LMC-DEB}. However, they are smaller by about 6-7\% than the original distance reported by \cite{SZEWCZYK-LMC}. As expected, our distances based on PL and PLZ relations are in very good agreement. Figure \ref{fig:PL_LMC} presents fits of intercepts of PL relations for the LMC based on $K_T$.

\begin{table*}
    \centering
    \begin{tabular}{|c|c||c|c|c|}
    \hline
   band & population & $\mu_{\rm LMC}$ & $\delta_{\rm LMC}$ & rms \\
        &       &   [mag] & [mag] &[mag] \\
   \hline
   \hline
                & RRab+RRc & 18.439 & 0.020 & 0.27\\
                \cline{2-5}
    $J$         & RRab & 18.450 & 0.027 & 0.28\\
                \cline{2-5}
                & RRc & 18.389 & 0.041 & 0.21\\
    \hline
    \hline
                & RRab+RRc & 18.438 & 0.020 & 0.22\\
                \cline{2-5}
    $K$         & RRab & 18.450 & 0.021 & 0.23\\
                \cline{2-5}
                &  RRc & 18.381 & 0.040  & 0.17\\
    \hline
    \hline
                & RRab+RRc & 18.431 & 0.019 & 0.20\\
                \cline{2-5}
    $K_T$         & RRab & 18.444 & 0.021 & 0.21\\
                \cline{2-5}
                &  RRc & 18.370 & 0.041  & 0.16\\
    \hline
    \hline
                & RRab+RRc & 18.439 & 0.019 & 0.24\\
                \cline{2-5}
    $W_{JK}$    & RRab & 18.451 & 0.019 & 0.25\\
                \cline{2-5}
                & RRc & 18.376 & 0.042 & 0.18\\
    \hline
    \end{tabular}
    \caption{LMC distance moduli $\mu_{\rm LMC} \pm \delta_{\rm LMC}$ obtained by fitting the zero point of a relation while keeping the period slopes fixed to calibrations presented in this work. The uncertainty $\delta_{\rm LMC}$ is a square root of the variance defined in Equation \ref{eq:covmu}. The rms of residuals of the fit is presented in the last column.}
    \label{tab:PL_SZEWCZYK}

    \centering
    \begin{tabular}{|c|c||c|c|c|}
    \hline
   band & population & $\mu_{\rm LMC}$ & $\delta_{\rm LMC}$ & rms \\
        &       &   [mag] & [mag] &[mag] \\
   \hline
   \hline
                & RRab+RRc & 18.450 & 0.020 & 0.26\\
                \cline{2-5}
    $J$         & RRab & 18.468 & 0.021 & 0.28\\
    \hline
    \hline
                & RRab+RRc & 18.445 & 0.016 & 0.21\\
                \cline{2-5}
    $K$         & RRab & 18.460 & 0.017 & 0.22\\
    \hline
    \hline
                & RRab+RRc & 18.438 & 0.016 & 0.20\\
                \cline{2-5}
    $K_T$       & RRab & 18.454 & 0.017 & 0.21\\
    \hline
    \hline
                & RRab+RRc & 18.441 & 0.015 & 0.24\\
                \cline{2-5}
    $W_{JK}$    & RRab & 18.454 & 0.016 & 0.25\\
    \hline
    \end{tabular}
    \caption{Same as Table~\ref{tab:PL_SZEWCZYK} but based on PLZ relations with the metallicity slope also set to a fixed value.}
    \label{tab:PLZ_SZEWCZYK}
\end{table*}

\subsection{Shifts of log P for RRc stars}

\begin{figure*}
    \centering
    \includegraphics[width=\textwidth]{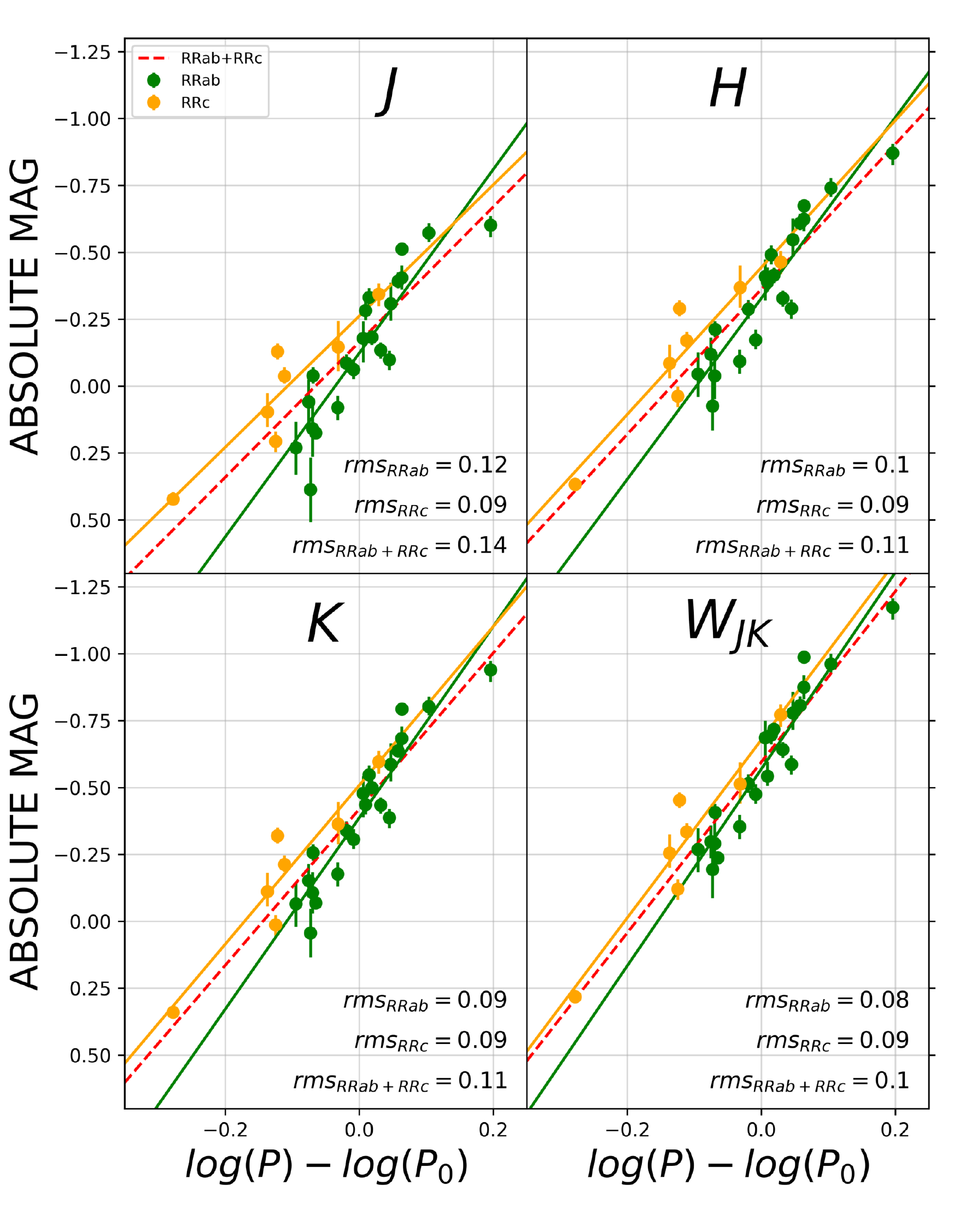}
    \caption{PL relations where only the fundamentalization of periods of RRc stars is applied ($\Delta \log P=0.127$). The suboptimal alignment of points corresponding to the two types of pulsators is caused by a selection effect.}
    \label{fig:PL127}
\end{figure*}

In one of the previous subsections, we compared the zero points of our calibrations with zero points of calibrations available in the literature. PLZ relations that are usually present in the literature are based on the $K_s-$ band and the mixed population (RRab+RRc) and thus require an application of $\Delta \log P$ for RRc stars. All quoted literature calibrations are based on $\Delta \log P =+0.127$, while we utilize our own $\Delta \log P$ values presented earlier in this work.

As we can notice in Figure \ref{fig:PL127}, or e.g. the work of (\citealt{VMC}; their Figures~6 \& 7), relations for the mixed population have significantly smaller slopes compared to relations for the two populations separately when $\Delta \log P=0.127$ is applied. RRc stars, even with properly fundamentalized periods, do not follow the same relations as RRab. They are still more luminous, which flattens the \textit{global} relation  for the mixed population and slightly increases its intercept. It again shows that a suboptimal alignment of points is generated by using the two populations of pulsators and can introduce a bias. 

RRc stars also constitute different fractions of different samples of mixed populations. When it comes to PLZ relations, the ratio of RRc to RRab stars in our sample -- $5/18$ ($28\%$) -- is slightly different from that of \cite{VMC}, which is $\sim 5,000/13,000$ ($38\%$) for $JK_s$ bands. In the Galactic sample used by \cite{MURAVEVA2015} this fraction is $2/21$ ($10\%$) and in \cite{MURAVEVA_GDR2} it is $35/366$ ($10\%$). In the case of the work of \cite{NEELEY}, the corresponding ratio is $17/38$ ($44\%$). The distributions of periods of RRc and RRab stars also vary from sample to sample. Thus, the weight and impact of the first-overtone pulsators on the final values of fitted parameters also differ between different samples. In principle, it makes the global relations containing a relatively large number of  first-overtone pulsators non-comparable (especially when it comes to values of individual parameters) with those having just a few.

Indeed, $\Delta \log P$ determined using RRd stars, which allow for the determination of ratios of the first-overtone to fundamental pulsation periods of these stars, yields values that have relatively small spread and do not differ substantively between different stellar systems (e.g. Petersen diagrams for RRd stars from the LMC and the SMC shown in \citealt{SOSZYNSKI}). They correspond well to the +0.127 shift. 

However, when it comes to the formation of common PL and PLZ relations for RRab+RRc stars, we should take into account that RRc stars populate a different region of HR diagrams than the fundamental-mode pulsators and double-mode pulsators (see e.g. \citealt{SZABO} and their Figure 5). RRc stars are hotter than RRab stars despite having the same luminosities. They have smaller radii and higher densities than RRab stars of the same luminosity. Thus their pulsation periods corresponding to the fundamental mode are shorter. Beyond the question of 'fundamentalization' of RRc stars periods, we are dealing here with a selection effect that should be corrected for in order to obtain PL and PLZ relations for the mixed population of RRab+RRc stars. In principle, such a correction is influenced by the different effective temperatures of different subsamples and can depend on wavelength; thereby, it should be determined independently for different photometric bands. When this is not possible, it is better to use PL(Z) relations just for RRL stars of one pulsating mode. Even though formally the uncertainty of the intercept decreases when including stars pulsating in both modes, the rms around the fit is increased and both the slope and intercept values might be biased in such a case.

\subsection{The influence of the parallax zero point}

A proper calibration of the Gaia EDR3 parallaxes is pivotal for any calibration of distance indicators. Unfortunately, this is a complex issue as the parallax zero point depends on several different variables such as stellar color, magnitude, or position of an object in the celestial sphere. The original Gaia EDR3 parallaxes used for the purpose of calibrations presented in this work were corrected using the \cite{LINDEGREN} values. The same corrections were used in the \cite{B-J2021} study in order to derive geometric and photo-geometric distances. They were estimated for individual objects based on distant quasars and LMC sources which provided a fixed reference frame. However, for bright stars ($G<13$\,mag, where $G$ is the Gaia photometric band) the authors had to use binaries with similar parallaxes but different colors and magnitudes. Note that all of the RRL stars from our sample have $G\sim9-11.5$\,mag. Given such physical pairs and using known biases for fainter sources, they were able to estimate biases for brighter companions and subsequently obtained a correcting relation for bright sources.
The median value of corrections for stars from our sample is $-0.029$\,mas\footnote{Such negative corrections were subtracted from the original Gaia EDR3 parallaxes, so that parallaxes became larger and the resulting distances smaller.}.

\begin{figure*}
    \centering
    \includegraphics[width=\textwidth]{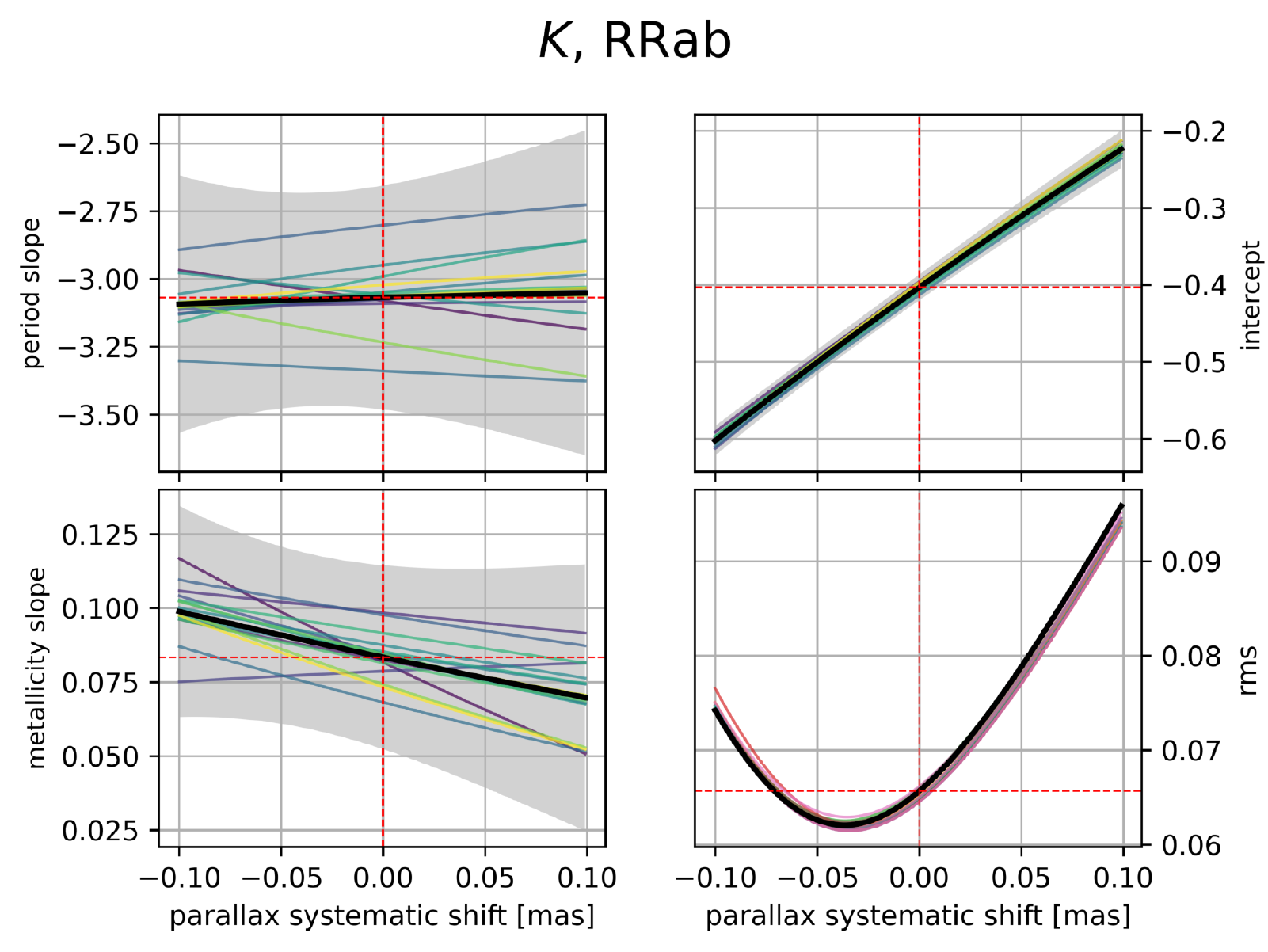}
    \caption{Dependence of PLZ parameters on the systematic parallax shift relative to parallaxes corrected according to \cite{LINDEGREN}. See the text for the explanation of lines.}
    \label{fig:K-dependence}
\end{figure*}

Figure~\ref{fig:K-dependence}, inspired by a similar analysis found in \cite{NEELEY}, depicts a relationship between parallax systematic shift, relative to parallaxes corrected using the \cite{LINDEGREN} corrections, and parameters of PLZ relation (using $K_s-$band and the RRab population as an example). Thick black lines denote fits using the whole RRab sample. Thinner colored lines correspond to jackknife resampling, where fits were performed for subsamples with different individual stars being rejected from the original sample. The gray zone denotes the $1 \sigma$ uncertainty of each parameter. The red dashed lines correspond to the values of parameters realized in the fit using the \cite{LINDEGREN} parallax corrections and the whole RRab sample. One can clearly see that the rms of the fit has a minimum - it generally depends on the band and the population of RRLs. This feature has been already reported by \cite{NEELEY} for their PLZ relations based on Gaia DR2 parallaxes. However, there are no reasons we should prefer a value of the shift that corresponds to a minimal rms. 

\begin{figure*}
    \centering
    \includegraphics[width=\textwidth]{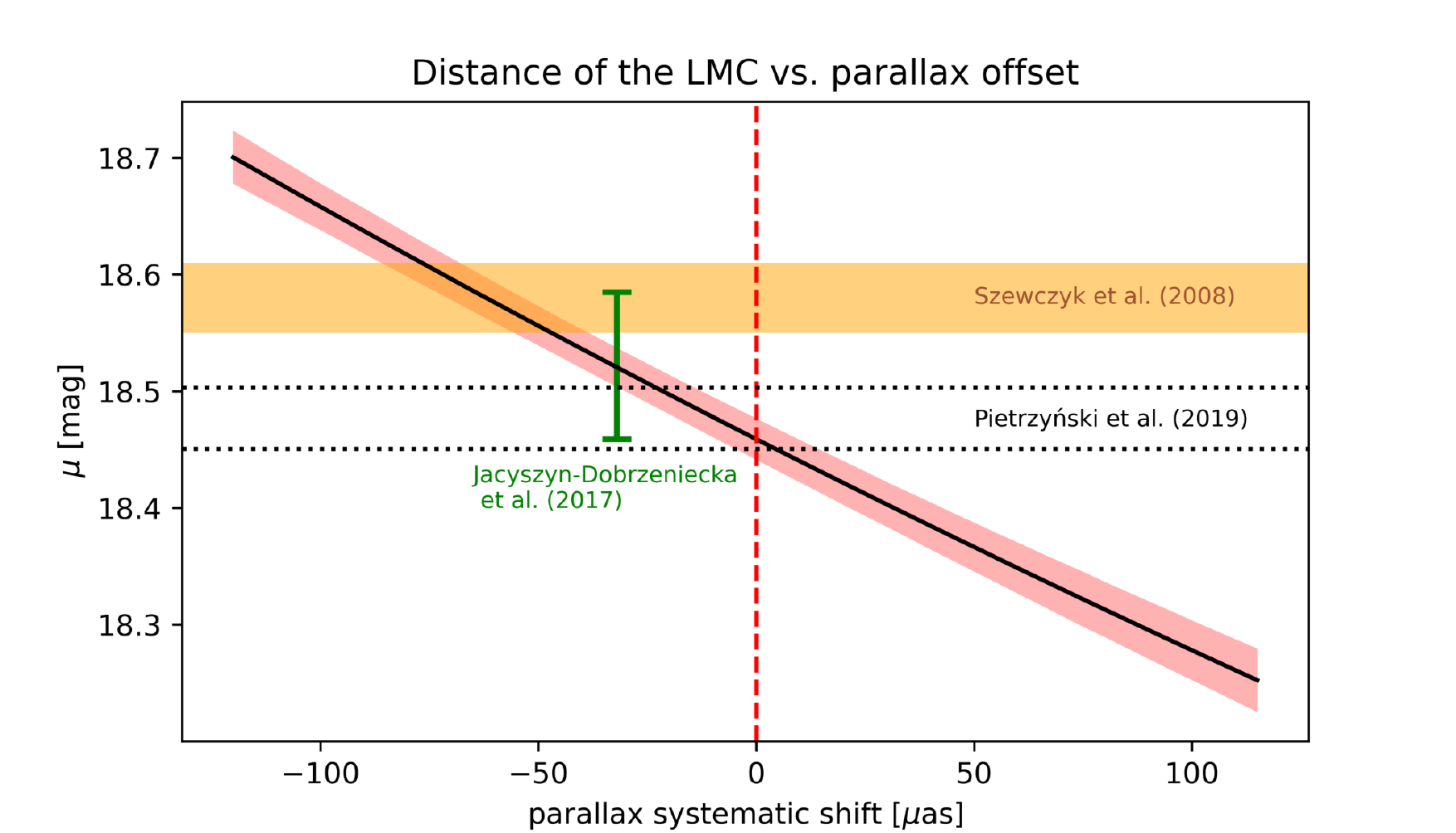}
    \caption{The dependence of the LMC distance ($K_s-$ band, RRab stars) on the systematic parallax offset, relative to GAIA EDR3 parallaxes with \cite{LINDEGREN} corrections, based on photometry of RR~Lyrae stars from the LMC given in \cite{SZEWCZYK-LMC}. Original distances reported in the literature are shown within their uncertainties for the reference.}
    \label{fig:LMC_OFFSET}
\end{figure*}

Figure~\ref{fig:LMC_OFFSET} presents the relation between parallax offset and distance to the LMC ($K_s-$band, fundamental pulsators only) obtained from the photometry of RRL stars in that galaxy published by \cite{SZEWCZYK-LMC}. The figure contains comparison of the offset-dependent distance based on our calibration with the canonical distance obtained using eclipsing binaries \citep{LMC-DEB} within its $1\sigma$ total error and the original distance (within its statistical uncertainty) reported by \cite{SZEWCZYK-LMC} that was based on calibrations of \cite{Bono2003}, \cite{CATELAN-PLZ}, and \cite{SOLLIMA}. The Figure also depicts the median distance of RR Lyrae stars reported by \cite{JACYSZYN} together with its total error.
The pink zone around the relation corresponds to the statistical uncertainty of the determined distance calculated using Equation \ref{eq:covmu}. The intercept of the red dashed line and the black solid line highlights the value of the LMC distance modulus obtained from our calibration. Thus we obtain an independent confirmation of the very close agreement between our zero point and the zero point anchored to the LMC distance obtained from eclipsing binaries.

\cite{LINDEGREN} estimate an uncertainty of a few $\mu as$ (microarcsec) for the parallax corrections. After assuming a conservative error of 10\,$\mu as$, we obtain a parallax component of the systematic uncertainty of the zero point of our calibration of around $0.02$\,mag.
This is the dominant component of the total systematic uncertainty for the zero point of PL and PLZ relations.

\section{Summary}

We have established new PL and PLZ relations for Galactic RR~Lyrae stars based on their NIR photometry from IRIS at Cerro Armazones Observatory and Gaia EDR3 parallaxes. We used 4 different methods to determine the relations, and obtained statistically identical results. The zero point of our relations is in very good agreement with the zero point based on RRL stars in the LMC and the distance to the Cloud obtained from eclipsing binaries given in \cite{LMC-DEB}. Three independent checks have confirmed this fact: fitting relations with period and metallicity slopes of \cite{VMC} and of \cite{MURAVEVA2015} (both of which are based on the VMC photometry) to our data, and by using our fiducial relations to derive a distance based on \cite{SZEWCZYK-LMC} photometry of RR~Lyrae stars from the LMC.
The zero point of our relations also agree very well with the calibration based on Baade-Wesselink distances to Galactic RR Lyrae stars \citep{MURAVEVA2015} even though those are of rather low precision.

Larger discrepancies are observed when comparing our calibrations with calibrations based on HST parallaxes of nearby RR Lyrae stars (\citealt{Sollima2006}, \citeyear{SOLLIMA}; \citealt{MURAVEVA2015}). Such zero points are not only based on parallaxes of lower precision, but are also based on one or a few stars. Given large uncertainties, they are still in agreement with our zero point. We also obtain a large difference in zero points when comparing with the \cite{NEELEY} relation based on Gaia DR2 parallaxes. However, \cite{MURAVEVA_GDR2} report relations that are also based on Gaia DR2 with different zero point of parallaxes and are in good agreement with our results.

In general, distances to the LMC obtained using PL and PLZ relations for $JK_s$ bands and the $W_{JK}$ index are in good agreement within their statistical errors. This is in part because of the small metallicity influence on NIR absolute magnitudes of $~0.1$\,mag/dex, similar to what has been reported before by, e.g. \cite{Sollima2006}, \cite{MURAVEVA2015}, and \cite{VMC}. 

The comparison of different calibrations for mixed populations of RRab+RRc stars is vague and may introduce a zero point and distance bias of a few hundredths of a magnitude compared to determinations based on one type of pulsators.
We presented calibrations for the mixed populations of the first-overtone and fundamental pulsators that include independent determinations of the logarithmic period shifts required for RRc stars. Our determinations are characterized by better alignment of the two types of pulsators along one relation because our method accounts for additional selection effects that may have biased studies where RRc stars had their periods fundamentalized only. 

Using the dependence of the LMC distance modulus on the parallax systematic offset, we estimated the systematic uncertainty of the zero point of our PL and PLZ relations as $0.02$\,mag. The combination of that uncertainty with a conservative estimation of the metallicity error (0.25 dex which corresponds to $\sim0.025$\,mag ) yields a systematic uncertainty of $0.03$\,mag for the LMC distance based on our calibrations of PLZ relations for RR Lyrae stars.

The research leading to these results has received funding from the European Research Council (ERC) under the European Union’s Horizon 2020 research and innovation program (grant
agreements No. 695099 \& 951549). 

The National Science Center (NCN) financed this research through
MAESTRO grant (agreement number UMO-2017/26/A/ST9/00446) and BEETHOVEN grant
(agreement number UMO-2018/31/G/ST9/03050). 

The research was possible thanks to
the grant of the Polish Ministry of Science and Higher Education (decision number DIR/WK/
2018/09).

WG gratefully acknowledges support from the ANID BASAL project ACE210002.

This research has made use of the International Variable Star Index (VSX) database, operated at AAVSO, Cambridge, Massachusetts, USA.

%

\appendix
\section{Covariances between fitted parameters of PL and PLZ relations}
\begin{table*}[h]
\centering
    \begin{tabular}{|c|c||c|c|c|}
        \hline
        \mc{2}{|l||}{PL relations} & \mc{3}{|c|}{photo-geometric distance \citep{B-J2021}}\\
        \hline
        \mc{1}{|l|}{band}& population & \mc{1}{|c|}{$cov(a,b) \times 10^5$} & \mc{1}{c|}{$cov(a,\Delta \log P)  \times 10^5$} & \mc{1}{c|}{$cov(b,\Delta \log P) \times 10^5$} \\
        \hline
        \hline
        & RRab+RRc & $-46$  & $-107$ & $16$ \\
        \cline{2-5}
        $J$ & RRab & $-106$  & $-$ & $-$ \\
        \cline{2-5}
	& RRc & $808$ & $-$ & $-$ \\
        \hline
        \hline
        & RRab+RRc & $-72$  & $-153$ & $18$ \\
        \cline{2-5}
        $H$ & RRab & $-122$  & $-$ & $-$ \\
        \cline{2-5}
	& RRc & $822$ & $-$ & $-$ \\
        \hline
        \hline
        & RRab+RRc & $-46$  & $-134$ & $15$ \\
        \cline{2-5}
        $K_s$ & RRab & $-66$  & $-$ & $-$ \\
        \cline{2-5}
	& RRc & $796$ & $-$ & $-$ \\
        \hline
        \hline
        & RRab+RRc & $-46$  & $-147$ & $15$ \\
        \cline{2-5}
        $W_{JK}$ & RRab & $-55$  & $-$ & $-$ \\
        \cline{2-5}
	& RRc & $820$ & $-$ & $-$ \\
        \hline
        \hline
	\hline
	\hline
	\mc{5}{|l||}{PLZ relations}\\
        \hline
        \mc{1}{|l|}{band}& population & \mc{1}{|c|}{$cov(a,b) \times 10^5$} & \mc{1}{c|}{$cov(a,c)  \times 10^5$} & \mc{1}{c|}{$cov(b,c) \times 10^5$} \\
        \hline
        \hline
        & RRab+RRc & $151$  & $629$ & $14$ \\
        \cline{2-5}
        $J$ & RRab & $26$  & $1350$ & $5$ \\
        \cline{2-5}
        \hline
        \hline
        & RRab+RRc & $157$  & $472$ & $18$ \\
        \cline{2-5}
        $H$ & RRab & $19$  & $1107$ & $1$ \\
        \cline{2-5}
        \hline
        \hline
        & RRab+RRc & $108$  & $357$ & $9$ \\
        \cline{2-5}
        $K_s$ & RRab & $11$  & $787$ & $3$ \\
        \cline{2-5}
        \hline
        \hline
        & RRab+RRc & $108$  & $313$ & $9$ \\
        \cline{2-5}
        $W_{JK}$ & RRab & $19$  & $752$ & $3$ \\
        \hline
        \hline
    \end{tabular}\par
    \caption{Covariances of fitted parameters of PL and PLZ relations based on the photo-geometric distances of \citep{B-J2021}; $a$ is a slope, $b$ - an intercept, $\Delta \log P$ is a fitted shift of $\log P$ for the first-overtone RRL when considering the mixed population of RRab+RRc, and $c$ is a metallicity slope in the case of PLZ relations.}
\label{tab:covs}
\end{table*}

\newpage
\section{Near-Infrared light curves of 28 nearby RR Lyrae stars}
\begin{figure}[h]
    \centering
    \includegraphics[width=0.85\textwidth]{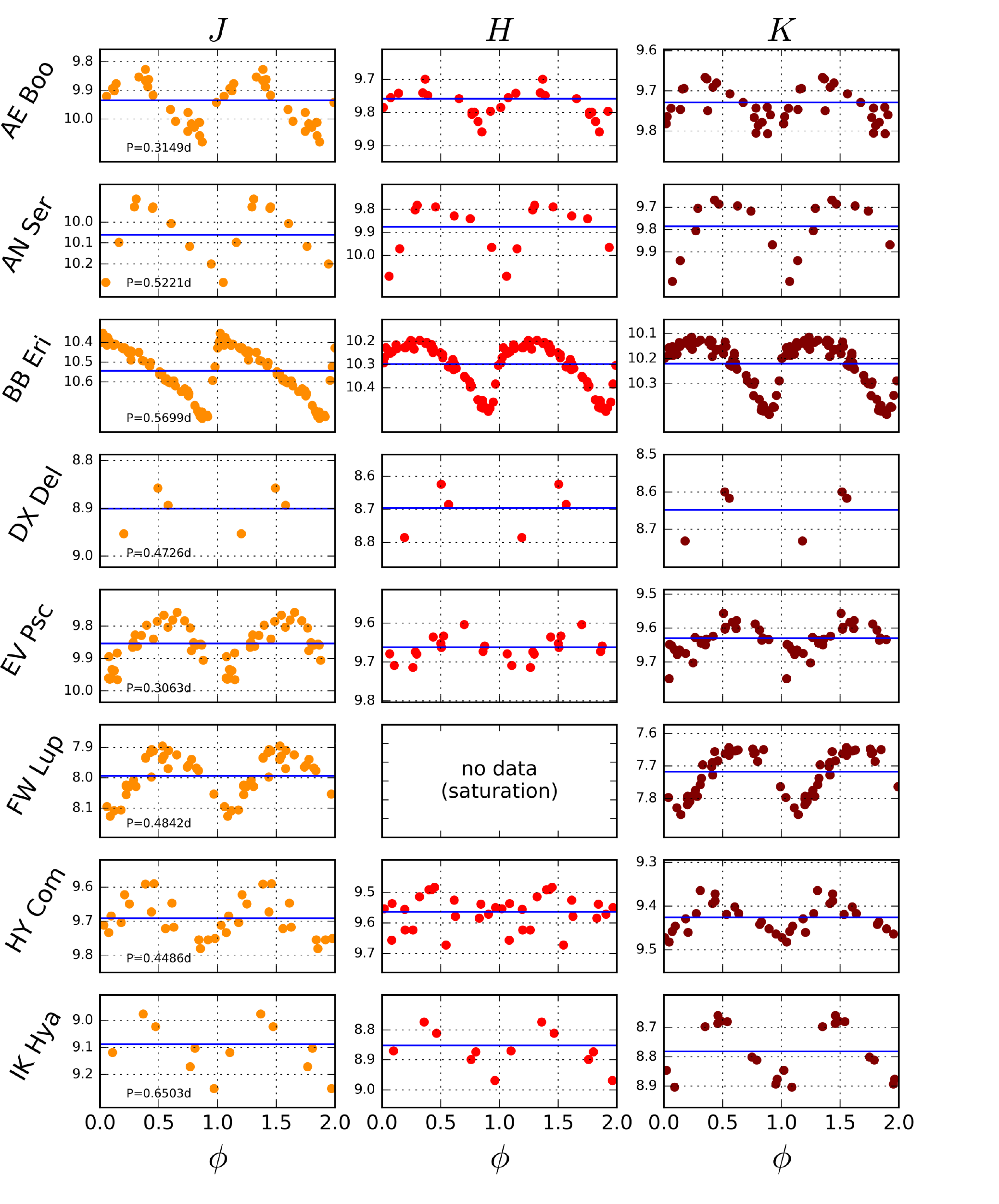}
    \caption{NIR light curves of stars used in this work. Horizontal blue lines correspond to determined mean magnitudes.}

\end{figure}
\begin{figure}
    \ContinuedFloat
    \centering
    \includegraphics[width=0.85\textwidth]{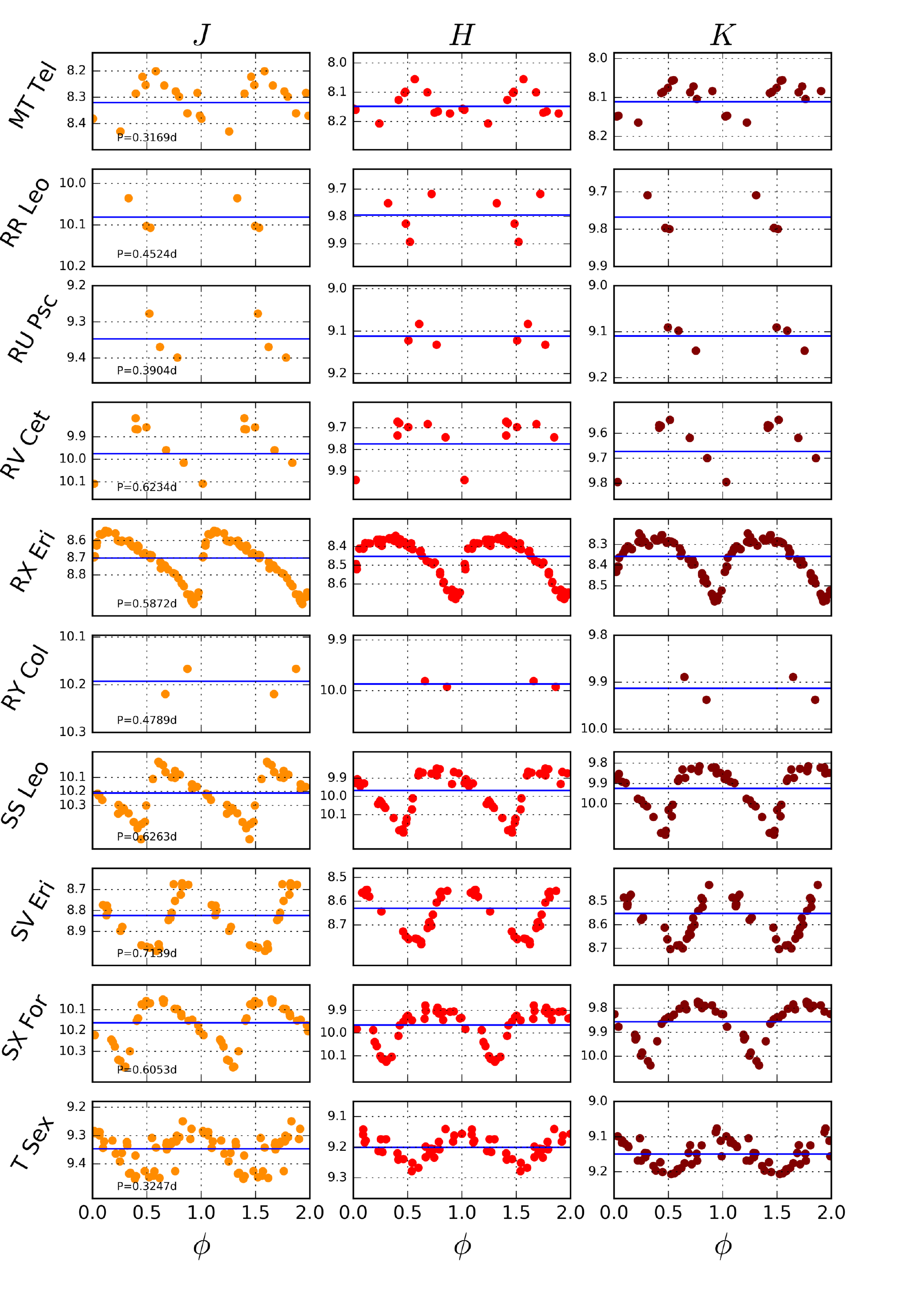}
    \caption{Continued}

\end{figure}
\begin{figure}
    \ContinuedFloat
    \centering
    \includegraphics[width=0.85\textwidth]{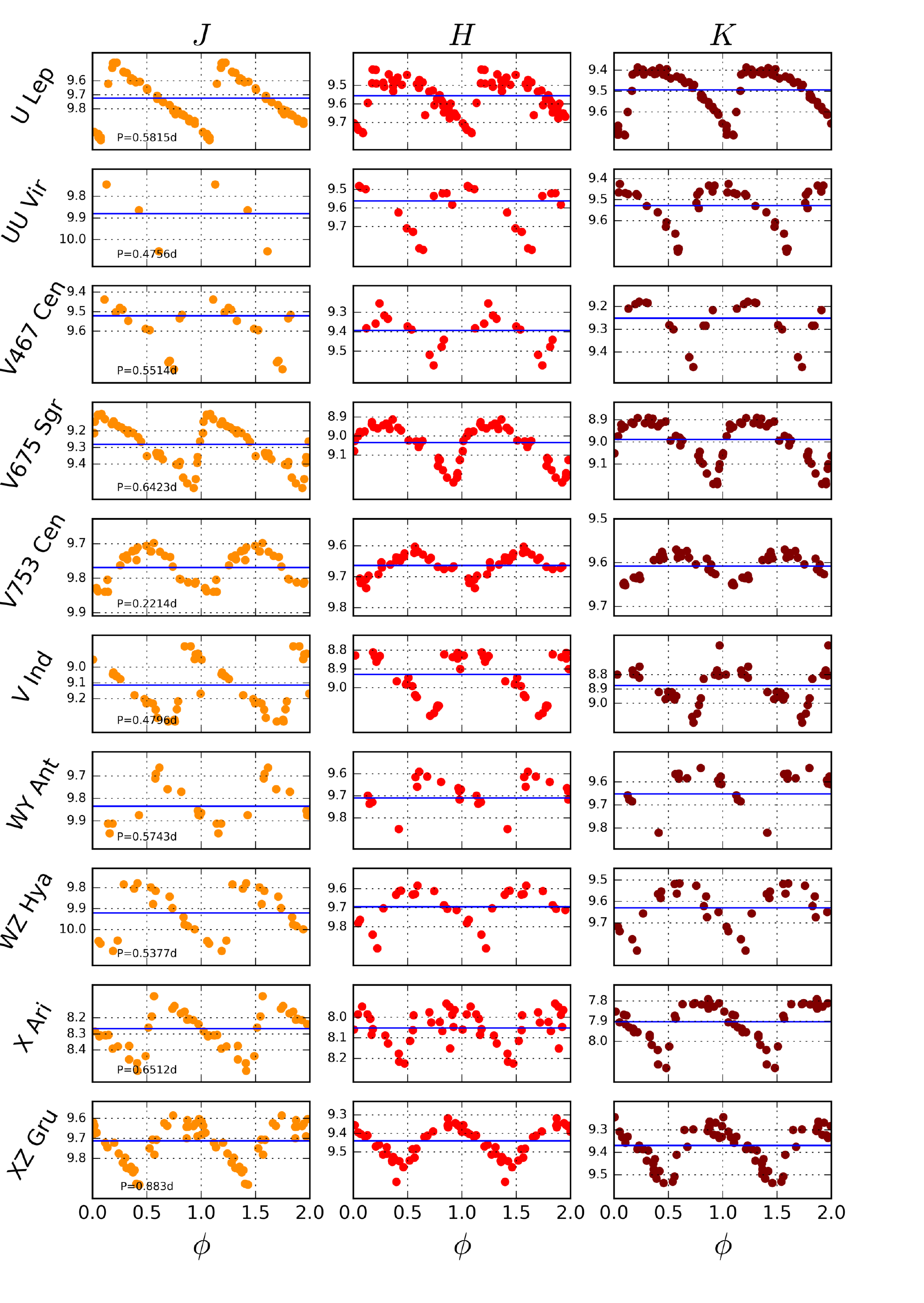}
    \caption{Continued}

\end{figure}
\end{document}